\documentclass[aps,pra,epsf,superscriptaddress,amsmath,amssymb,amsfonts,twocolumn,showpacs]{revtex4-2}

\usepackage{graphicx} 
\usepackage[caption=false]{subfig} 
\usepackage{dcolumn}
\usepackage{pict2e}
\usepackage{multirow} 
\usepackage{booktabs}
\usepackage{color} 
\usepackage[dvipsnames]{xcolor} 
\usepackage{amsmath, bm} 
\usepackage{amsfonts, amssymb}  
\usepackage{verbatim}
\usepackage{underscore} 
\newcommand{\ket}[1]{\ensuremath{\vert#1\rangle}}  
\newcommand{\bra}[1]{\ensuremath{\langle#1\vert}} 
\newcommand{\braketoperator}[3]{\ensuremath{\left\langle#1\left\lvert#2\right\rvert#3\right\rangle}} 

\newcommand{\ion}[2]{#1\ensuremath{^{#2+}}}

\begin{document}

\title{Theoretical investigation of orbital alignment of x-ray-ionized atoms in exotic electronic configurations}

\author{Laura Budewig}
\affiliation{Center for Free-Electron Laser Science CFEL, Deutsches Elektronen-Synchrotron DESY, Notkestrasse 85, 22607 Hamburg, Germany}\affiliation{Department of Physics, Universit\"at Hamburg, Notkestrasse 9-11, 22607 Hamburg, Germany} 
\author{Sang-Kil Son}
\affiliation{Center for Free-Electron Laser Science CFEL, Deutsches Elektronen-Synchrotron DESY, Notkestrasse 85, 22607 Hamburg, Germany}
\affiliation{The Hamburg Centre for Ultrafast Imaging, Luruper Chaussee 149, 22761 Hamburg, Germany}
\author{Robin Santra}
\affiliation{Center for Free-Electron Laser Science CFEL, Deutsches Elektronen-Synchrotron DESY, Notkestrasse 85, 22607 Hamburg, Germany}\affiliation{Department of Physics, Universit\"at Hamburg, Notkestrasse 9-11, 22607 Hamburg, Germany}
\affiliation{The Hamburg Centre for Ultrafast Imaging, Luruper Chaussee 149, 22761 Hamburg, Germany}

\date{\today}

\begin{abstract}
We theoretically study orbital alignment in x-ray-ionized atoms and ions, based on improved electronic-structure calculations starting from the Hartree-Fock-Slater model. 
We employ first-order many-body perturbation theory to improve the Hartree-Fock-Slater calculations and show that the use of first-order-corrected energies yields significantly better transition energies than originally obtained.
The improved electronic-structure calculations enable us also to compute individual state-to-state cross sections and transition rates and, thus, to investigate orbital alignment induced by linearly polarized x rays. 
To explore the orbital alignment of transiently formed ions after photoionization, we discuss  alignment parameters and ratios of individual state-resolved photoionization cross sections for initially neutral argon and two exotic electronic configurations that may be formed during x-ray multiphoton ionization dynamics induced by x-ray free-electron lasers. We also present how the orbital alignment is affected by Auger-Meitner decay and demonstrate how it evolves during a sequence of one photoionization and one Auger-Meitner decay. 
Our present work establishes a step toward investigation of orbital alignment in atomic ionization driven by high-intensity x rays.
\end{abstract}

\maketitle 

\section{Introduction}\label{Introduction}
The development of x-ray free-electron lasers (XFELs) around the world~\cite{LCLS,SACLA,PAL-XFEL,EuXFEL,SwissXFEL} enables scientists to study a variety of new fields in structural biology~\cite{Spence, Schlichting, Coe, Chapman2}, ultrafast x-ray atomic and molecular physics~\cite{R, Bostedt, Fang, Ullrich}, as well as dense matter physics~\cite{Glenzer}, owing to ultraintense and ultrashort x-ray radiation with an unprecedentedly high brilliance~\cite{Pellegrini2}.
Prototype examples of the excellent opportunities of XFELs are serial femtosecond crystallography~\cite{Chapman} and single-particle imaging experiments~\cite{Sobolev,Seibert},
which permit structure determination with almost atomic resolution~\cite{Branden, Hosseinizadeh, Assalauova, Aquila, Ziaja}. 

Accurate theoretical simulations  in combination with experimental studies~\cite{Fukuzawa, Rudek, Rudek2, Young2, Doumy, Rudenko} have shown that extremely highly ionized atomic ions can be produced during interaction with ultraintense and ultrashort x-ray pulses. 
In general, an x-ray photon is absorbed by an inner-shell electron, which is followed by a decay process via Auger-Meitner decay or x-ray fluorescence~\cite{Cixp}.
Further photoionization with accompanying decay cascades lead to very highly charged states of atoms~\cite{Fukuzawa, Rudek, Rudek2, Young2, Doumy} or molecules~\cite{Rudenko}, which is called x-ray multiphoton ionization~\cite{IoIXBwA}. 
In the molecular case,  the sample undergoes structural disintegration, which limits the resolution achievable in x-ray imaging experiments~\cite{Nass, Quiney, Lorenz, Son}.

On the other hand, it has been well known for a long time that ions produced by single photoionization commonly exhibit an alignment~\cite{Flugge, Jacobs, Caldwell, Greene2} due to different ionization probabilities of ions with different projection quantum numbers. A theoretical treatment of alignment, including a description of parameters to quantify the alignment and orientation, can be found in Refs.~\cite{Blum, Greene, Schmidt}. 
More recently, orbital alignment has, for instance, been explored for single photoionization of initially closed-shell atoms and cations with respect to spin-orbit coupling~\cite{Kleiman2, Kleiman} and 
for strong-field ionized atoms~\cite{Young, SFCoXP}.

As a consequence of these aspects, orbital alignment during x-ray multiphoton ionization becomes a topic of interest, which remains  relatively unexplored. It addresses the orbital alignment of the highly charged ions produced in the end of multiple sequences of photoabsorption and accompanying relaxation events, but also  how the alignment changes during the x-ray multiphoton ionization dynamics. A first critical step in this research direction is to develop a suitable atomic structure framework that provides individual \(LS\) eigenstates as well as individual state-to-state cross sections and  transition rates for each angular momentum projection~\(M_L\). In a later step this framework can then be embedded in an ionization dynamics calculation, so that individual states can also be captured during x-ray multiphoton ionization dynamics.

In this work, we present such a first step by extending the \emph{ab initio} electronic-structure toolkit \textsc{xatom}~\cite{X,Son}. \textsc{xatom} is a useful and successful tool~\cite{Doumy, Rudek2, Rudek, Fukuzawa} for simulating x-ray-induced atomic processes and ionization dynamics of neutral atoms, atomic ions, and ions embedded in a plasma~\cite{Thiele2, Thiele}.
Based on a Hartree-Fock-Slater (HFS) calculation of orbitals and orbital energies, subshell photoionization cross sections and fluorescence and Auger-Meitner group rates~\cite{Son} can be computed among other things~\cite{Bekx2, Slowik, Son2017, Toyota}.
These quantities are employed to determine the ionization dynamics  by solving a set of coupled rate equations~\cite{IoIXBwA} either directly~\cite{Son} or via more efficient Monte Carlo algorithms for heavy atoms~\cite{Son2012, Fukuzawa}. 
Since especially for heavy atoms an extremely huge number of electronic configurations are involved in the ionization dynamics,  computational efficiency is critical. For instance, for xenon atoms this number can be estimated to be \(\sim\)2.6\(\times10^{68}\) when relativistic and resonant effects are included \cite{Rudek}.  Therefore, the \textsc{xatom} toolkit uses HFS, one of the simplest and most efficient first-principles electronic-structure methods.
Even though there are other more accurate atomic structure toolkits (see e.g.,~\cite{Fischer}), improving the \textsc{xatom} toolkit is of critical relevance for investigations in x-ray-induced ionization dynamics. Computational efficiency becomes even more crucial for solving state-resolved rate equations.  In this case the number of individual electronic states involved in the ionization dynamics goes far beyond the number of involved electronic configurations.

We extend the \textsc{xatom} toolkit~\cite{XATOM, X, Son} by incorporating an improved electronic-structure description, based on the first-order many-body perturbation theory. This permits the computation of first-order-corrected energies and a set of zeroth-order $LS$ eigenstates, for arbitrary electronic configurations. Moreover, individual state-to-state photoionization cross sections and transition rates are calculated, based on our new implementation. 
A detailed comparison with experimental results available in the literature shows that the extended \textsc{xatom} toolkit delivers significantly improved transition energies in contrast to the original version.  
The knowledge of individual state-to-state cross sections enables us to study orbital alignment of ions produced by single photoionization. We focus on the orbital alignment of transiently formed ions resulting from photoionization of a neutral argon atom and some exotic electronic configurations of argon by linearly polarized x rays. 
An important remark here is that these ions can be viewed as examples of species appearing in x-ray multiphoton ionization of neutral atoms. It would be possible to observe them experimentally with ultrafast XFEL pulses by using a transient absorption experiment with two-color x rays, similar to that employed in Ref.~\cite{Young}.  
Using the individual state-to-state transition rates, we investigate any change of orbital alignment after fluorescence and Auger-Meitner decay. Combining state-to-state cross sections and rates, we examine the orbital alignment after one x-ray-induced ionization process, i.e., a sequence comprising a photoionization event and a subsequent relaxation event.
Understanding of the orbital alignment of this x-ray single-photon ionization will be a building block to explore and explain the orbital alignment occurring during x-ray multiphoton ionization dynamics induced by interaction with intense XFEL pulses.

The paper is organized as follows. 
In Sec.~\ref{theory} we briefly present the theoretical framework of our implementation in \textsc{xatom} and discuss quantities to quantify the alignment. The validation of our implementation is the topic of Sec.~\ref{validation}. 
Orbital alignment in initially neutral Ar, \ion{Ar}{} ($2p^{-1}$), and \ion{Ar}{2} ($2p^{-2}$) is studied in Sec.~\ref{results}. We also briefly address the orbital alignment caused by x-ray-induced ionization including relaxation in this section.
We conclude with a summary and future perspectives in Sec.~\ref{conclusion}.  

\section{Theoretical details}\label{theory} 
The aim of this section is to outline the theoretical framework. 
In particular, we start with the HFS Hamiltonian, whose solutions are already present in the \textsc{xatom} toolkit~\cite{XATOM}. 
Then we develop a method to determine first-order-corrected energies and a new set of eigenstates by employing first-order degenerate perturbation theory. 
These improved electronic-structure calculations are implemented as an extension of the \textsc{xatom} toolkit~\cite{XATOM} and are further utilized to calculate individual state-to-state cross sections and transition rates.
Throughout this paper, atomic units, i.e., \(m=\vert e\vert=\hbar=1\) and \(c=1/\alpha\), are used, where \(\alpha\) is the fine-structure constant.

\subsection{The Hamiltonian}\label{Hamiltonian}
The Hamiltonian describing \(N\) nonrelativistic electrons in an atom can be separated into the HFS Hamiltonian, \({\hat{H}}^{\text{HFS}}\), and the residual electron-electron interaction, \({\hat{V}}_{\text{res}}\), defined by the full two-electron interactions minus the HFS mean field [see Eq.~(\ref{eqn:Vres})],
\begin{equation}
\label{eqn:Hmatter}
{\hat{H}}_{\text{matter}} ={\hat{H}}^{\text{HFS}}+{\hat{V}}_{\text{res}}.
\end{equation}
The one-electron solutions of the HFS Hamiltonian are so-called spin orbitals \({\varphi}_{q}\) and spin orbital energies \({\varepsilon}_{q}\), respectively~\cite{Cixp}. 
Consequently, we solve the following effective one-electron equation:
\begin{equation}
\label{eqn:one-electron-H}
\left[ -\frac{1}{2}\nabla^2+\hat{V}^{\text{HFS}}(\vec{x})\right]{\varphi}_{q}(\vec{x})={\varepsilon}_{q}{\varphi}_{q}(\vec{x}).
\end{equation}
Here, \(\hat{V}^{\text{HFS}}\) is the Hartree-Fock-Slater mean field~\cite{HFS},
\begin{equation}
\label{eqn:VHFS}
\hat{V}^{\text{HFS}}(\vec{x}) = - \frac{Z}{\vert\vec{x}\vert}+
\int d^3x^{\prime}\frac{\rho({\vec{x}}^\prime)}{\vert{\vec{x}}-{\vec{x}}^\prime\vert} -\frac{3}{2}\left[\frac{3}{\pi}\rho(\vec{x})\right]^{\frac{1}{3}},
\end{equation}
with the local electron density \(\rho(\vec{x})\) and the nuclear charge \(Z\).
A more extensive description of how to numerically solve Eq.~(\ref{eqn:one-electron-H}) within the \textsc{xatom} toolkit can be found in, e.g., Refs.~\cite{X, Son}. 
In the context of this paper it is only worth mentioning that  the spin orbitals can be decomposed into a radial part, a spherical harmonic and a spin part~\cite{Son},
\begin{equation}
{\varphi}_{q}(\vec{x}) = \frac{{u}_{{\xi}_{q},{l}_{q}}(r)}{r}{{Y}}_{{l}_{q}}^{{m}_{{l}_{q}}}(\Omega)\begin{pmatrix}
{\delta}_{{m}_{{s}_{q}},\frac{1}{2}}\\
{\delta}_{{m}_{{s}_{q}},-\frac{1}{2}}\end{pmatrix}.
\end{equation}
Accordingly, the index \(q\) refers to a set of four quantum numbers \(({\xi}_{q}, {l}_{q},{m}_{{l}_{q}},{m}_{{s}_{q}})\) (with \({\xi}_{q}={n}_{q}\) for bound spin orbitals, i.e., \({\varepsilon}_{q}<0\), and \({\xi}_{q}={\varepsilon}_{q}\) for unbound spin orbitals, i.e., \({\varepsilon}_{q}\geq0\)). 
Note that spin orbitals belonging to the same subshell, i.e., the same  \(n\) and \(l\) quantum numbers, share the same orbital energy, denoted as \({\varepsilon}_{nl}\).

Introducing anticommutating creation and annihilation operators, \({\hat{c}^\dagger}_{q}\) and \({\hat{c}}_{q}\), associated with the spin orbitals~\cite{Cixp,Gottfried:QM}, the two parts of Eq.~(\ref{eqn:Hmatter}) can be expressed as  
\begin{equation}
\label{eqn:HHFS}
{\hat{H}}^{\text{HFS}}=\sum_{q}{\varepsilon}_{q}{\hat{c}^\dagger}_{q}{\hat{c}}_{q}
\end{equation}
and 
\begin{equation}
\label{eqn:Vres}
{\hat{V}}_{\text{res}}=-\sum_{p,q}{V}_{pq}^{\text{HFS}}{\hat{c}^\dagger}_{p}{\hat{c}}_{q}+\frac{1}{2}\sum_{p,q,r,s}{v}_{pqrs}{\hat{c}^\dagger}_{p}{\hat{c}^\dagger}_{q}{\hat{c}}_{s}{\hat{c}}_{r}.
\end{equation}
In these expressions the summations run over all spin orbitals. Furthermore, 
\begin{equation}
{V}_{pq}^{\text{HFS}}=\int d^{3}x\,{\varphi}_{p}^\dagger(\vec{x})\hat{V}^{\text{\text{HFS}}}(\vec{x}){\varphi}_{q}(\vec{x})
\end{equation}
 is a mean-field matrix element and 
 \begin{equation}
 \label{eqn:vpqrs}
 {v}_{pqrs}=\int \! \int d^{3}x\,d^{3}{x'}\,{\varphi}_{p}^\dagger(\vec{x}){\varphi}_{q}^\dagger(\vec{x'})\frac{1}{\vert \vec{x}-\vec{x'}\vert }{\varphi}_{r}(\vec{x}){\varphi}_{s}(\vec{x'})
 \end{equation}
is a two-electron Coulomb matrix element.
 
Having at hand the one-electron eigenstates the $N$-electron eigenstates of \(\hat{H}^{\text{HFS}}\) are then formed by an antisymmetrized product~\cite{Gottfried:QM}. 
Known as electronic Fock states, these multi-orbital states read
\begin{equation}
\label{eqn:FS}
\ket{{\Phi}_{\alpha}} = \Biggr\vert \prod_{q=1}^\infty {n}_{q}^{\alpha} \Biggr\rangle = \prod_{q=1}^\infty({\hat{c}^\dagger}_{q})^{{n}_{q}^{\alpha}}\ket{0},
\end{equation}
where \(\ket{0}\) is the vacuum and the occupation number  \({n}_{q}^{\alpha} \in \lbrace 0,1 \rbrace\)  is  restricted by \(\sum_{q=1}^{\infty}{n}_{q}^{\alpha}=N\).
The energy of a Fock state is
\begin{equation}
\label{eqn:EFS}
{E}_{\alpha} = \sum_{q=1}^{\infty}{n}_{q}^{\alpha}{\varepsilon}_{q} = \sum_{n,l} {N}_{nl}{\varepsilon}_{nl},
\end{equation}
with the latter summation running over all subshells, occupied by \({N}_{nl}\) electrons. Note that the Fock states are only eigenstates of \({\hat{H}}^{\text{HFS}}\) but not of \({\hat{H}}_{\text{matter}}\).

\subsection{Improved electronic-structure calculations}\label{DPT}
In order to obtain  approximate solutions of \({\hat{H}}_{\text{matter}}\), we employ  first-order time-independent degenerate perturbation theory~\cite{Griffiths:QM,Sakurai:QM}. 
Regarding \({\hat{H}}_{\text{matter}}\) in Eq.~(\ref{eqn:Hmatter}), \({\hat{H}}^{\text{HFS}}\) is treated as the unperturbed Hamiltonian, the  well-known Fock states \(\ket{{\Phi}_{\alpha}}\) as the unperturbed zeroth-order states, and \({\hat{V}}_{\text{res}}\) as the perturbation. 
Since the Fock states belonging to the same electronic configuration share the same energy with respect to  \({\hat{H}}^{\text{HFS}}\) [see Eq.~(\ref{eqn:EFS})], degenerate perturbation theory has to be applied.

Here the method implemented to determine the first-order-corrected energy eigenvalues of \({\hat{H}}_{\text{matter}}\) and a new set of zeroth-order eigenstates by employing first-order degenerate perturbation theory is briefly sketched. 
In particular, we assume that an arbitrary electronic configuration is given for the atom or ion for which we want to find the solutions.  
Then an application of degenerate perturbation theory requires the following steps.

\begin{enumerate}
\item \textbf{Find the set of Fock states belonging to the given electronic configuration and make subsets according to ($M_S$, $M_L$).} 
It is useful to group the Fock states \(\ket{{\Phi}_{\alpha}}\) into subsets according  to the total spin projection \({M}_{S}^{\alpha} = \sum_{q}{n}_{q}^{\alpha}{m}_{{s}_{q}}\) and the projection of the total angular momentum operator \({M}_{L}^{\alpha} = \sum_{q}{n}_{q}^{\alpha}{m}_{{l}_{q}}\). 
Therefore, from now on a Fock state is expressed as \(\ket{{\Phi}_{\gamma}^{{M}_{S};{M}_{L}}}\), with its projection quantum numbers as upper labels and with a lower index $\gamma$ that runs from 1 to the number of Fock states with $M_S$ and $M_L$. 
Then for each subset \(\lbrace\ket{{\Phi}_{\gamma}^{{M}_{S};{M}_{L}}}\rbrace\) the Fock states are separately determined as strings of occupation numbers, i.e., zeros and ones [see Eq.~(\ref{eqn:FS})].

\item \textbf{Compute the  matrix elements of \({\bm{H}}_{\text{matter}}\) within each subset and diagonalize each submatrix \({\bm{H}}_{\text{matter}}^{\bold{({M}_{S},{M}_{L})}}\).} 
Most importantly, utilizing the Condon  rules~\cite{Condon:TAS}, it can be easily shown that \({\bm{H}}_{\text{matter}}\) is block diagonal in the previously introduced subsets  \(\lbrace\ket{{\Phi}_{\gamma}^{{M}_{S};{M}_{L}}}\rbrace\). 
Therefore, it is sufficient to compute only matrix elements within the subsets, i.e., \(\braketoperator{{\Phi}_{\delta}^{{M}_{S};{M}_{L}}}{{\hat{H}}_{\text{matter}}}{{\Phi}_{\gamma}^{{M}_{S};{M}_{L}}}\), and to numerically diagonalize each submatrix \({\bm{H}}_{\textmd{matter}}^{({M}_{S},{M}_{L})}\) separately. 
The eigenvalues of \({\bm{H}}_{\textmd{matter}}^{({M}_{S},{M}_{L})}\) deliver first-order-corrected energies and its eigenstates offer a new subset of zeroth-order states having projection quantum numbers \({M}_{S}\) and \({M}_{L}\). 
These new states are linear combinations of the Fock states \(\ket{{\Phi}_{\gamma}^{{M}_{S};{M}_{L}}}\) belonging to the subset in question. 
In contrast to the Fock states the new states have the advantage of being also eigenstates of total orbital angular momentum and of total spin. 
Thus, from now on we refer to the new zeroth-order states as zeroth-order $LS$ eigenstates. 

\item \textbf{Identify the term symbol for each pair of first-order-corrected energy and zeroth-order $\bm{LS}$ eigenstate.}
To label the zeroth-order $LS$ eigenstates we use the set of quantum numbers \((L,S,{M}_{L},{M}_{S})\) together with an additional integer index $\kappa$ that runs from 1 to the number of states with $(L,S,{M}_{L},{M}_{S})$. 
Consequently, the zeroth-order $LS$ eigenstates read
\begin{equation}
\label{eqn:newstateLS}
\ket{LS{M}_{L}{M}_{S}\kappa} = \sum_{\gamma}{c}_{LS{M}_{L}{M}_{S}\kappa}^{\gamma}\ket{{\Phi}_{\gamma}^{{M}_{S};{M}_{L}}},
\end{equation}  
with the expansion coefficient \({c}_{LS{M}_{L}{M}_{S}\kappa}^{\gamma}\) obtained from step 2. 
The values for the projection quantum numbers are directly known from the subset in question, whereas the other labels, i.e., \(L\), \(S\), and \(\kappa\), need to be determined. 
The zeroth-order $LS$ eigenstates, having the same values for \( L\), \(S\), and \(\kappa\), form a term~\cite{Alonso:QM}, which is characterized by a term symbol \(^{2S+1}L\left(\kappa\right)\). 
Also note that all states within a term share the same energy. 
So let the first-order-corrected energies be denoted by \({E}_{LS\kappa}\).  
Combining this knowledge with the method of Slater diagrams, described in, e.g., Refs.~\cite{Alonso:QM,Struve:MS}, \(L\), \(S\), and \(\kappa\) can be identified for each pair of first-order-corrected energy and zeroth-order $LS$ eigenstate. 

\item \textbf{If terms share the same first-order-corrected energy, diagonalize \(\bm{{S}}^2\) and/or \(\bm{{L}}^2\) with respect to the zeroth-order $\bm{LS}$ eigenstates in question.} 
This step is necessary to guarantee that the zeroth-order states are all proper $LS$ eigenstates in the end (for more details see~\cite{Condon:TAS}). 
\end{enumerate} 

The method described delivers all terms \(^{2S+1}L\left(\kappa\right)\) together with their first-order-corrected energies \({E}_{LS\kappa}\) and all zeroth-order $LS$ eigenstates \(\ket{LS{M}_{L}{M}_{S}\kappa}\) for a given atom or ion in a given electronic configuration. 
In the following for simplicity the label \(\kappa\) is omitted, either because \(\kappa=1\) for all involved states or because it is irrelevant in the computation in question.
We point out that the orbitals and their energies needed to create the submatrices \({\bm{H}}_{\textmd{matter}}^{({M}_{S},{M}_{L})}\) are provided by the original \textsc{xatom} toolkit.
Moreover, note that there exists an alternative way of constructing the \(LS\) eigenstates by employing Racah algebra~\cite{Racah1, Racah2, Racah3, Judd:OTiAS}. However, here we have used the strategy that was developed by \citet{Condon:TAS}. Numerical diagonalization of the relatively small matrices that arise in the approach we adopt  does not determine the overall computational effort.

\subsection{Individual state-to-state photoionization cross sections}\label{CS}
Having at hand first-order-corrected energies and zeroth-order $LS$ eigenstates [Eq.~(\ref{eqn:newstateLS})] for the initial and final electronic configurations, we can compute photoionization cross sections for these individual initial and final states. 
Note that from now on the index \(i\) appears when referring to the quantities of the initial state, i.e., \(\ket{{L}_{i}{S}_{i}{M}_{{L}_{i}}{M}_{{S}_{i}}}\), while the index \(f\) is used for a final target state, i.e., \(\ket{{L}_{f}{S}_{f}{M}_{{L}_{f}}{M}_{{S}_{f}}}\). We remark that the final target state excludes the unbound photoelectron. Thus, the total final electronic state is given by \(\ket{{L}_{f}{S}_{f}{M}_{{L}_{f}}{M}_{{S}_{f}};{\varepsilon}_{c}{l}_{c}{m}_{{l}_{c}}{m}_{{s}_{c}}}\), where the latter quantum numbers refer to those of the photoelectron. However, attributed to the use of a fully uncoupled approach for the continuum states, the total final state is no $LS$ eigenstate.
In order to calculate the cross section for orthogonal spin orbitals and for photons linearly polarized along the \(z\) axis~\cite{Hertel:AMOP, X-rayspectroscopy}, we use the first-order time-dependent perturbation theory~\cite{Cixp} and the electric dipole approximation~\cite{IoIXBwA}. As long as we integrate over the photoelectron angular distribution, the latter approximation, utilized also in the original \textsc{xatom} toolkit~\cite{Son}, works well. 
Then the individual state-to-state cross section for ionizing  an electron in the subshell with quantum numbers \(n\) and \(l\) by absorbing a linearly polarized photon with  energy \({\omega}_{\text{in}}\) may be written as
\begin{widetext}
\begin{equation}
\label{eqn:pcs}
\begin{split}
{\sigma}_{^{2S_i+1}L_i;^{2S_f+1}L_f}^{{M}_{{L}_{i}};{M}_{{L}_{f}}}(nl,{\omega}_{\text{in}})
=&\frac{4\pi^2}{3{\omega}_{\text{in}}}\alpha{({\varepsilon}_{nl}-{\varepsilon}_{c})^2}\sum_{{l}_{c}=\left\vert{l}-1\right\vert}^{{l}+1}{l}_{>}\left\vert\int_{0}^{\infty}dr{u}_{{\varepsilon}_{c} {l}_{c}}^{\ast}(r)r{u}_{{n} {l}}(r)\right\vert^2\\
&\times\sum_{{M}_{{S}_{f}}}\left\vert C({l},{l}_{c},1;{M}_{{L}_{i}}-{M}_{{L}_{f}},{M}_{{L}_{f}}-{M}_{{L}_{i}},0)\braketoperator{{L}_{f}{S}_{f}{M}_{{L}_{f}}{M}_{{S}_{f}}}{{\hat{c}}_{j}}{{L}_{i}{S}_{i}{M}_{{L}_{i}}{S}_{i}}\right\vert^2.
\end{split}
\end{equation}
\end{widetext}
In this expression, \(C(\cdots)\) represents a Clebsch-Gordan coefficient~\cite{Racah2, Judd:OTiAS}, \({l}_{>} = \text{max}({{l}_{c},{l}})\), \({\varepsilon}_{c}={\omega}_{\text{in}}+{E}_{{L}_{i}{S}_{i}}-{E}_{{L}_{f}{S}_{f}}\) is the energy of the photoelectron, and \({\varepsilon}_{nl}\) is the orbital energy of the \(nl\) subshell \footnote{ Due to the inclusion of first-order correction in the energy,  \({E}_{{L}_{i}{S}_{i}}-{E}_{{L}_{f}{S}_{f}}\) differs from \({\varepsilon}_{nl}\). Consequently, the cross section contains a prefactor \({({\varepsilon}_{nl}-{\varepsilon}_{c})^2}/{{\omega}_{\text{in}}}\) rather than the usual prefactor \({\omega}_{\text{in}}\).}. Owing to the selection rules for a dipole transition~\cite{Hertel:AMOP},  the first sum in Eq.~(\ref{eqn:pcs}) does not include \({l}_{c}={l}\).
Following the independent-particle model~\cite{Crasemann:ISP}, the index \(j\) of the involved bound spin orbital (i.e., from which an electron is ejected) refers to the set of quantum numbers \(({n}_{j},{l}_{j},{m}_{{l}_{j}},{m}_{{s}_{j}}) = ({n},{l},{M}_{{L}_{i}}-{M}_{{L}_{f}},{M}_{{S}_{i}}-{M}_{{S}_{f}})\) with the restriction \({M}_{{S}_{i}}={S}_{i}\). 
The interaction Hamiltonian causing one-photon absorption~\cite{Cixp} does not affect the spin and its projection. 
Thereby, the cross section is independent of the initial spin projection \({M}_{{S}_{i}}\) when performing a summation over the final spin projection \({M}_{{S}_{f}}\). 
Accordingly, Eq.~(\ref{eqn:pcs}) describes a transition between one initial zeroth-order $LS$ eigenstate \(\ket{{L}_{i}{S}_{i}{M}_{{L}_{i}}{M}_{{S}_{i}}}\) with arbitrary spin projection \({M}_{{S}_{i}}\) and both final zeroth-order $LS$ eigenstates \(\ket{{L}_{f}{S}_{f}{M}_{{L}_{f}}{M}_{{S}_{i}}+\frac{1}{2}}\) and \(\ket{{L}_{f}{S}_{f}{M}_{{L}_{f}}{M}_{{S}_{i}}-\frac{1}{2}}\), as long as \(\vert{M}_{{S}_{i}}\pm\frac{1}{2}\vert\le{S}_{f}\). 
We also remark that the matrix element \(\braketoperator{{L}_{f}{S}_{f}{M}_{{L}_{f}}{M}_{{S}_{f}}}{{\hat{c}}_{j}}{{L}_{i}{S}_{i}{M}_{{L}_{i}}{S}_{i}}\) can be interpreted as the overlap between the final and initial zeroth-order $LS$ eigenstates. Due to the involved annihilation operator \({\hat{c}}_{j}\) (see Sec.~\ref{Hamiltonian}), the initial state is, however, already reduced by the involved electron in the spin orbital \({\varphi}_{j}\). 
The larger (smaller) this overlap is, the larger (smaller) the cross section is. 
If the overlap is zero, the cross section is zero.

\subsection{Individual state-to-state transition rates}\label{TR}
The fluorescence rate for a transition of an electron from the \(n_j l_j\) subshell  to a hole in the lower lying  \(n_h l_h\) subshell can be calculated in a similar way as the photoionization cross section~\cite{Cixp, Hertel:AMOP,IoIXBwA}.  
Accordingly, the individual state-to-state fluorescence rate  associated with a transition from the initial zeroth-order $LS$ eigenstate \(\ket{{L}_{i}{S}_{i}{M}_{{L}_{i}}{M}_{{S}_{i}}}\)  to an accessible final target state \(\ket{{L}_{f}{S}_{f}{M}_{{L}_{f}}{M}_{{S}_{f}}}\) may be written as
\begin{widetext}
\begin{equation}
\label{eqn:ftr}
\begin{split}
{\Gamma}_{^{2S_i+1}L_i;^{2S_f+1}L_f}^{{M}_{{L}_{i}};{M}_{{L}_{f}}}(n_j l_j,n_h l_h)
=&\frac{4l_>}{3(2l_h+1)}\alpha^3({E}_{L_i S_i}-{E}_{L_f S_f}){({\varepsilon}_{n_h l_h}-{\varepsilon}_{n_j l_j})^2}\left\vert\int_{0}^{\infty}dr{u}_{n_h {l}_{h}}^{\ast}(r)r{u}_{{n_j} {l_j}}(r)\right\vert^2\\
&\times\left\vert \sum_{h,j}C(1,{l}_{j},l_h;{M}_{{L}_{f}}-{M}_{{L}_{i}},{m}_{{l}_{j}},{m}_{l_h})\braketoperator{{L}_{f}{S}_{f}{M}_{{L}_{f}}{{S}_{f}}}{{\hat{c}}_{h}^\dagger{\hat{c}}_{j}}{{L}_{i}{S}_{i}{M}_{{L}_{i}}{S}_{i}}\right\vert^2.
\end{split}
\end{equation}
\end{widetext}
Here, \({l}_{>} = \text{max}({{l}_{j},{l}_{h}})\). The indices \(j\) and \(h\) of the solely initially or respectively finally occupied bound spin orbitals (between which the electron is transferred) refer to the set of quantum numbers \(({n}_{j},{l}_{j},{m}_{{l}_{j}},{m}_{{s}_{j}}) \)  and \(({n_h},{l_h},{m}_{{l}_{h}}={M}_{L_f}-{M}_{L_i}+{m}_{l_j},{m}_{{s}_{h}}={m}_{s_j})\). We note that  the total  projections \({M}_{L_i}\) and \({M}_{L_f}\)  determine the relation between \({m}_{l_j}\) and \({m}_{l_h}\) only, but not their values. Consequently, it is necessary to include in Eq.~(\ref{eqn:ftr}) a summation over all possible spin orbitals that are only occupied in the initial or the final state, respectively.
Since the interaction Hamiltonian~\cite{Cixp} does not affect the spin and its projection and \({M}_{{S}_{i}}={M}_{{S}_{f}}\) due to the selection rules, the transition rate  is independent of the initial and final spin projection. 

Another transition rate of interest is the Auger-Meitner decay rate that two electrons in the  \(n_j l_j\) and \(n_{j^\prime} l_{j^\prime}\) subshells 
undergo transitions: one into a hole in the lower-lying  \(n_h l_h\) subshell 
 and the other into the continuum. Since this process occurs via the electron-electron interaction, its transition rate can be obtained by employing the first-order time-dependent perturbation theory with the interaction Hamiltonian given by Eq.~(\ref{eqn:Vres})~\cite{Cixp, X-rayspectroscopy}.
Accordingly, the individual state-to-state Auger-Meitner decay rate   may be written as
\begin{widetext}
\begin{equation}
\label{eqn:amtr}
{\Gamma}_{^{2S_i+1}L_i;^{2S_f+1}L_f}^{{M}_{{L}_{i}};{M}_{{L}_{f}}}(n_j l_j, n_{j^\prime} l_{j^\prime}, n_h l_h)
=2\pi\sum_{l_a}\sum_{M_{S_f}}\left\vert \sum_{h,j, j^\prime}\lbrack{v}_{ah j j^\prime}-{v}_{ah j^\prime j}\rbrack \braketoperator{{L}_{f}{S}_{f}{M}_{{L}_{f}}{M}_{{S}_{f}}}{{\hat{c}}_{h}^\dagger{\hat{c}}_{j^\prime}{\hat{c}}_{j}}{{L}_{i}{S}_{i}{M}_{{L}_{i}}{S}_{i}}\right\vert^2,
\end{equation}
\end{widetext}
where  \({v}_{ah j j^\prime}\) and \({v}_{ah j^\prime j}\) are both two-electron Coulomb matrix elements given by Eq.~(\ref{eqn:vpqrs}). Here, the index \(a\) refers to the quantum numbers of the Auger electron, i.e., \(({\varepsilon}_{a}={E}_{{L}_{i}{S}_{i}}-{E}_{{L}_{f}{S}_{f}},{l}_{a},{M}_{{L}_{i}}-{M}_{{L}_{f}},M_{{S}_{i}}-{M}_{{S}_{f}})\).  The sum over \(l_a\) is restricted by \(\vert{L}_{f}-{L}_{i}\vert\leq l_a\leq\text{min}(l_h+l_j+l_{j^\prime},{L}_{f}+{L}_{i})\). The indices \(j\), \(j^\prime\), and \(h\) denote the involved spin orbitals in the corresponding \(n_j l_j\), \(n_{j^\prime} l_{j^\prime}\), and \(n_h l_h\) subshells, respectively.  However, it is necessary to include in Eq.~(\ref{eqn:amtr}) a summation over all possible spin orbitals that are either only occupied in the initial or the final state, because the projection quantum numbers of these orbitals are not completely definite.  Additionally, including a summation over the final spin projection \({M}_{S_f}\) provides a transition rate that  is independent of the initial spin projection. 

\subsection{Alignment parameter and ratios}\label{A20}
We briefly introduce a few basic quantities that we employ for the investigation of orbital alignment.
The alignment parameter~\cite{Schmidt, Blum, Greene} offers a measure of the alignment of a final ion with definite angular momentum \(L_f\) due to different projections \({M}_{L_f}\). Recall that there is no coupling to the spin for the employed interaction Hamiltonians~\cite{Cixp} and, thus, no alignment with respect to \(M_{S_f}\). Therefore, in what follows we neglect the spin \(S_f\) and its projection \(M_{S_f}\), i.e., in this section \(\ket{L_f {M}_{L_f}}\) refers to a final zeroth-order $LS$ eigenstate, a sum over accessible \({M}_{S_f}\) included as in Eqs.~(\ref{eqn:pcs}) and (\ref{eqn:amtr}). Here, we briefly define the alignment parameter in our context. Further discussions and applications of the alignment and orientation parameters can be found in Refs.~\cite{Blum,Greene, Schmidt}.

Let us start with the density matrix for the \(L_f\) under investigation~\cite{Blum, Greene},
\begin{equation}
\hat{\rho}= \sum_{M_{L_f}}p({M}_{L_f}\vert L_f)\ket{L_f M_{L_f}}\bra{L_f {M}_{L_f}}.
\end{equation}
Here, \(p({M}_{L_f}\vert L_f)\) is the conditional population probability of  the final state with projection \({M}_{L_f}\) for a given \(L_f\). For single photoionization this probability is given by  \(p({M}_{L_f}\vert L_f)={\sigma}_{^{2S_i+1}L_i;^{2S_f+1}L_f}^{{M}_{{L}_{i}};{M}_{{L}_{f}}}/\sum_{{M}_{L_f}}{\sigma}_{^{2S_i+1}L_i;^{2S_f+1}L_f}^{{M}_{{L}_{i}};{M}_{{L}_{f}}}\) for an  \(S_f\) and an  initial state. Similarly the population probability can be obtained for the decay processes via the transition rates in Eqs.~(\ref{eqn:ftr}) and (\ref{eqn:amtr}).
As a next step, we decompose \(\hat{\rho}\) in terms of irreducible spherical tensor operators~\cite{Greene, Fano}
\begin{equation}
\begin{split}
{\hat{T}}_{JM}=\sum_{M_{L_f},M_{{L}_{f}}^\prime}(-1)^{L_f-M_{{L}_{f}}^\prime}&C(L_f,L_f,J;M_{L_f},-M_{{L}_{f}}^\prime,M)\\
\times &\ket{L_f M_{L_f}}\bra{L_f M_{{L}_{f}}^\prime}.
\end{split}
\end{equation}
Thus, we may write
\begin{equation}
\hat{\rho}= \sum_{J,M}\rho_{JM}{\hat{T}}_{JM},
\end{equation}
where the expansion coefficients \(\rho_{JM}\) are given by~\cite{Blum} 
\begin{equation}
\begin{split}
\rho_{JM}=\sum_{{M}_{L_f}}&(-1)^{L_f-{M}_{L_f}}p({M}_{L_f}\vert L_f)\\
&\times C(L_f,L_f,J;M_{L_f},-M_{L_f},M).
\end{split}
\end{equation}
Note that these coefficients are only nonzero for \(M=0\). In the case of \(p(-{M}_{L_f}\vert L_f)=p({M}_{L_f}\vert L_f)\), \(\rho_{J0}\) vanishes for odd \(J\)  owing to the properties of the Clebsch-Gordan coefficient~\cite{Racah2}. Since there is no preference for \(\pm M_{L_f}\) in the interaction with linearly polarized light, i.e., \(p(-{M}_{L_f}\vert L_f)=p({M}_{L_f}\vert L_f)\), \(\rho_{10}\) is always zero. Thus,  the orientation parameter, defined by \(\mathcal{O}_{10}=\rho_{10}/\rho_{00}\)~\cite{Schmidt}, is zero and no orientation is created by the interaction with linearly polarized light. 

The coefficients \(\rho_{J0}\) are also known as statistical tensors, and they define the alignment parameter as follows~\cite{Schmidt}:
\begin{equation}
\label{eqn:A20}
\begin{split}
\mathcal{A}_{20}(L_f) =&\rho_{20}/\rho_{00}\\
=& \sqrt{\frac{5}{(2L_f+3)(L_f+1)L_f(2L_f-1)}}\\
&\times \sum_{M_{L_f}}\left[3{M}_{L_f}^2-L_f(L_f+1)\right] p({M}_{L_f}\vert L_f).
\end{split}
\end{equation}
To obtain the second line of Eq.~(\ref{eqn:A20}) we have utilized the formula for the Clebsch-Gordan coefficients given  in Ref.~\cite{Racah2}.
Note that \(\mathcal{A}_{20}\) is positive (negative), when states with larger (smaller) \(\vert {M}_{L_f}\vert\) are more likely populated than the others. For a uniform distribution, \(\mathcal{A}_{20}=0\) (no alignment), while the larger \(\vert\mathcal{A}_{20}\vert\) the stronger the alignment. 

In general an ion produced by photoionization can have  different \(L_f\), but the alignment parameter can only capture one \(L_f\). Thus, if we are interested  in the distribution of all possible final states, then another quantity of interest is the ratio of individual (state-resolved) cross sections,
\begin{equation}
\label{eqn:ratio}
{\sigma}_{^{2S_i+1}L_i;^{2S_f+1}L_f}^{{M}_{{L}_{i}};{M}_{{L}_{f}}}(nl,{\omega}_{\text{in}})/{\sigma}_{^{2S_i+1}L_i}^{{M}_{{L}_{i}}}(nl,{\omega}_{\text{in}}).
\end{equation}
This provides direct information about the probability to find the ion produced  in the final \(^{2{S}_{f}+1}{L}_{f}\) state with projection \({M}_{{L}_{f}}\), when the atom or ion is initially in the \(^{2{S}_{i}+1}{L}_{i}\) state with projection \({M}_{{L}_{i}}\) and when the \(nl\) subshell  is ionized.
Here, \({\sigma}_{^{2S_i+1}L_i}^{{M}_{{L}_{i}}}(nl,{\omega}_{\text{in}})\) is the subshell cross section of the \(nl\) subshell  restricted to the initial state in question, summing over all possible final states.  
We do not distinguish between states with different spin projection as the cross sections are independent of it.  Moreover, replacing the cross sections in Eq.~(\ref{eqn:ratio}) by the corresponding transition rates [Eqs.~(\ref{eqn:ftr}) and (\ref{eqn:amtr})], delivers ratios of individual transition rates.

\section{Validation}\label{validation}
Having discussed the basic formalism underlying our implementation in \textsc{xatom}, we next proceed to explore transition energies and photoionization cross sections for explicit electronic configurations and to compare the results with  experimental measurements. 
In particular, we employ two different theoretical strategies for describing physical processes. 
In the zeroth-order strategy, transition energies are computed based on zeroth-order energies for the initial and final states. 
The zeroth-order energies are the sum of orbital energies according to the initial or final electronic configuration [Eq.~(\ref{eqn:EFS})]. 
On the other side, in the first-order strategy, transition energies are computed based on the first-order-corrected energies \({E}_{LS}\) for the initial and final states (see Sec.~\ref{DPT}). 
For both strategies, energies and radial integrals are calculated with orbitals and orbital energies optimized for the initial configuration only. 
The usage of the same set of orbitals for both the initial and final configurations avoids issues with orbital nonorthogonality~\cite{Olsen, Verbeek, Hayes}.
Moreover, it should be mentioned that we still perform zeroth-order calculations using the original version of \textsc{xatom}, whereas for the first-order calculations we employ the present implementation. 

Orbitals and orbital energies are numerically solved on a radial grid employed by \textsc{xatom} (see Refs.~\cite{Son, X} for details), based on  the HFS potential [Eq.~(\ref{eqn:VHFS})] including the Latter tail correction~\cite{Latter}. 
In what follows, the bound states are computed using the generalized pseudospectral method~\cite{Yao,Tong} on a nonuniform grid with 200 grid points and a maximum radius of 50~a.u. The continuum states are computed using the fourth-order Runge-Kutta method on a uniform grid~\cite{Cooper, Manson} with a grid size of 0.005~a.u., employing the  same potential as used in the bound-state calculation. It has been demonstrated that the cross sections and rates calculated using \textsc{xatom} (zeroth-order strategy) show good agreement with available experimental data and other calculations~\cite{Son, Toyota,Son2012}. 

\subsection{Transition energies for neon}
\begin{figure}[t]  
\centering
\includegraphics[width=\linewidth]{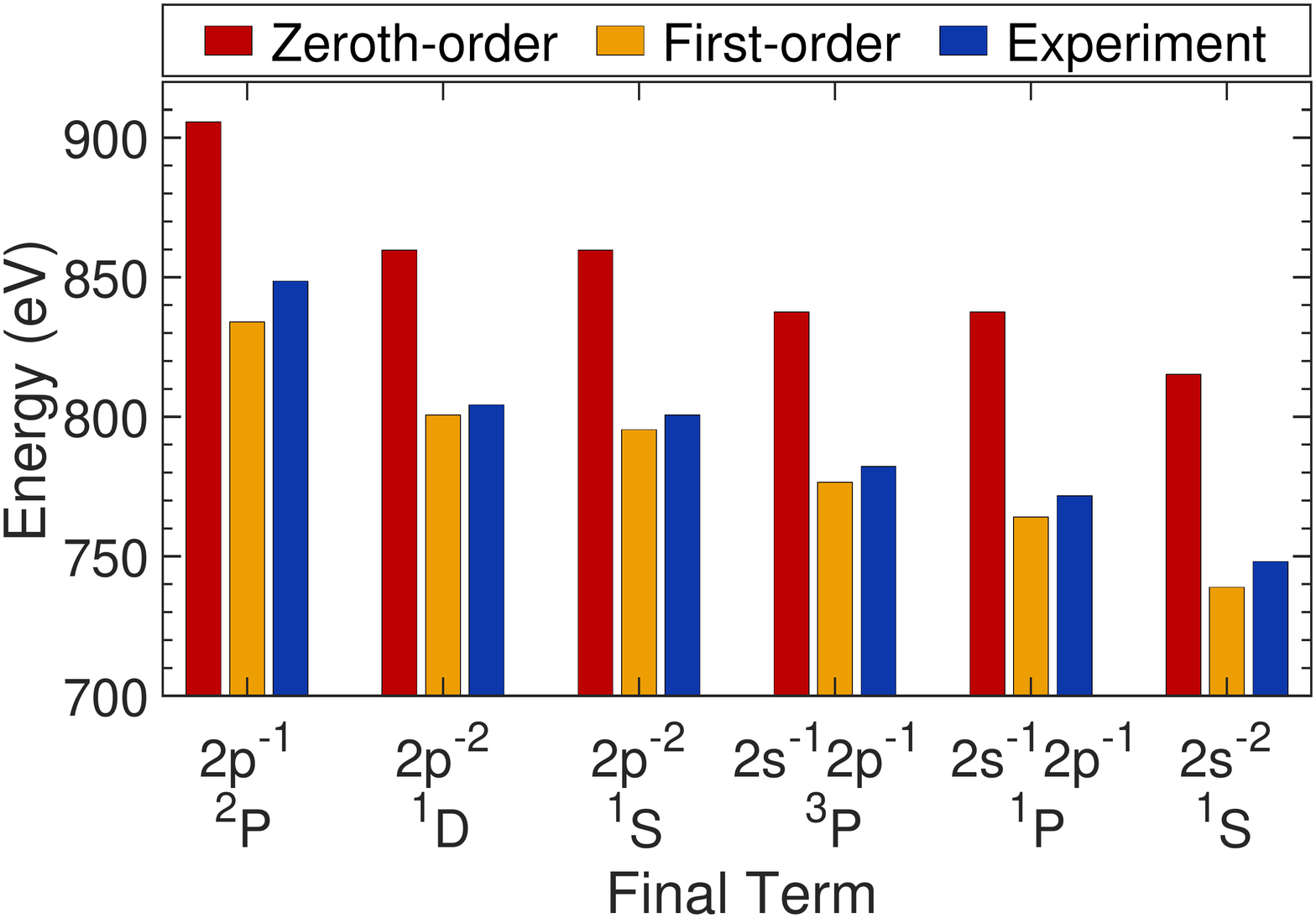}
\caption{Comparison of experimental \(K\alpha\) fluorescence energy \cite{Booklet, X-ray} and \(KLL\) Auger-electron energies \cite{Albiez} for neon with two theoretical strategies (see legend). The different lines are labeled by the final open subshell(s) (first line) and by the final term symbol (second line). 
In all cases the \ion{Ne}{} ion is initially in the $1s^{-1}$ $^2$S state.
}
\label{fig:energyNeon}
\end{figure}
First, the \(K\alpha\) fluorescence energy and all \(KLL\) Auger-electron energies are examined for an initial \ion{Ne}{} ion with a $K$-shell vacancy ($1s^{-1}$ $^2$S). 
The results are presented in Fig.~\ref{fig:energyNeon}. 
It is apparent from the data that the first-order strategy is in reasonable agreement with the experimental $K\alpha$ fluorescence energy~\cite{X-ray,Booklet} and the $KLL$ Auger-Meitner electron energies~\cite{Albiez}, to within less than \(2\%\).  
In contrast, the energies obtained via the zeroth-order strategy differ significantly from the experimental values.
These findings indicate that the first-order strategy, contained in our implementation, is the better strategy for describing the transition energy.
The small difference between experiment and theory still remaining for the first-order calculation might be attributed to the use of the same set of initial and final orbitals, the neglect of higher-order terms, and relativistic effects.   
We remark that no value is shown for the final state of $2p^{-2}$ $^3$P in  Fig.~\ref{fig:energyNeon}, because this transition is forbidden on account of parity~\cite{Cixp, Auger}.

\subsection{Photoionization cross sections for argon}
\begin{figure}[t]  
\centering
\subfloat{
\includegraphics[width=\linewidth]{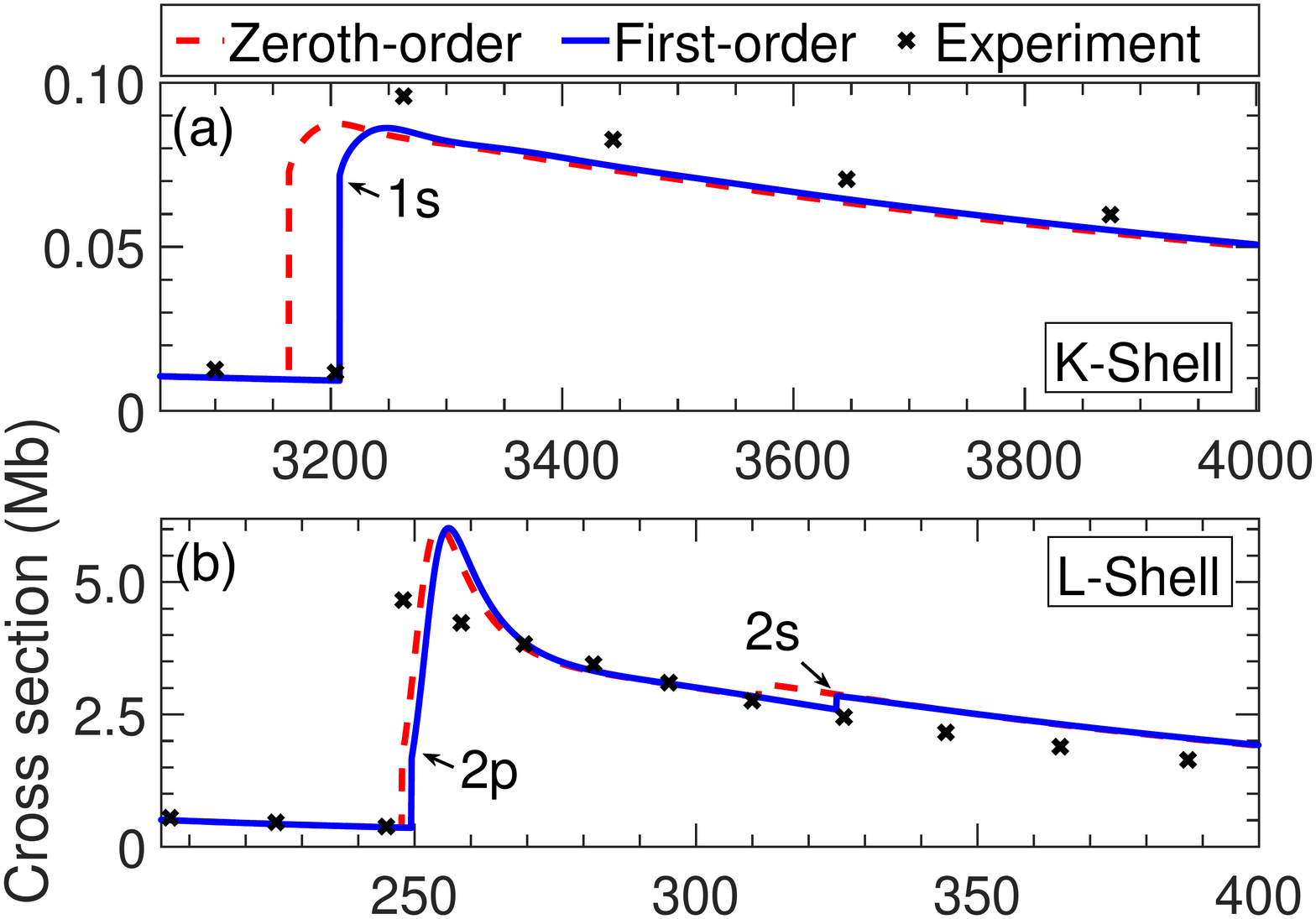}} \\[-0.9ex]
\subfloat{
\includegraphics[trim = 0mm 97.5mm 0mm 0mm,  clip,width=\linewidth]{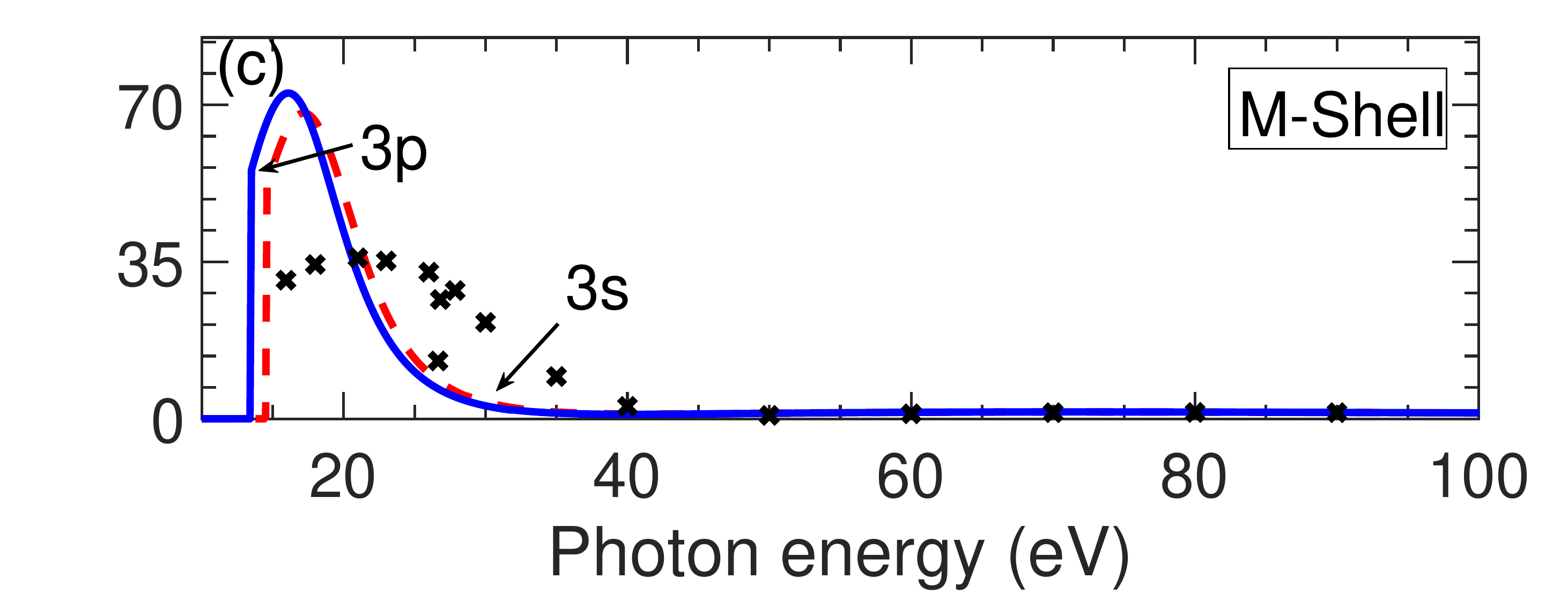}} 
\caption{Calculated total photoionization cross section, in Mb, of neutral argon as a function of the photon energy, in eV. Results for both the first-order strategy (solid blue line) and the zeroth-order strategy (dashed red line) are compared to experimental data (black crosses) reported in Ref.~\cite{Marr} for the $K$- and $L$-shell thresholds and Ref.~\cite{Samson} for the $M$-shell threshold.}
\label{fig:pcsAr}
\end{figure}
As a next example, we examine photoionization of a neutral argon atom (\(1s^22s^22p^63s^23p^6\)) in the region of the thresholds. 
Figure~\ref{fig:pcsAr} shows the total photoionization cross section as a function of the photon energy in the (a) \(K\)-shell, (b) \(L\)-shell, and (c) \(M\)-shell threshold regions. 
The total cross sections, which are an incoherent sum over all individual state-to-state cross sections in Eq.~(\ref{eqn:pcs}), are depicted. 
Interestingly, with regard to the cross section, both the first-order and zeroth-order strategies behave very similar, if one ignores the shift due to different threshold energies.
In general, both are in acceptable agreement with the experimental values around the $K$- and $L$-shell thresholds~\cite{Marr} and the $M$-shell threshold~\cite{Samson}, often to within less than \(10\%\). 
However, especially at the $L$- and $M$-shell thresholds, the calculated cross sections are significantly higher than the experimental values (more than \(50\%\) between the \(3p\) and \(3s\) thresholds and $\sim$25\% at the \(2p\) and \(2s\) thresholds). 
This observed disagreement and the lack of improvement concerning the first-order strategy mainly stem from the use of zeroth-order states in both strategies.
In the present framework, first-order-corrected energies but only zeroth-order eigenstates are calculated (see Sec.~\ref{DPT}). If the first-order states were used, it would be possible to capture a part of interchannel coupling~\cite{IC}, since the first-order states are a mixture of the zeroth-order states from different electronic configurations~\cite{Sakurai:QM}.
Moreover, it should be mentioned  that the experimental results in Fig.~\ref{fig:pcsAr}(c) contain resonances between the \(3p\) and the \(3s\) thresholds~\cite{Samson}, but  the theoretical calculations do not. 

Even though our implementation only leads to an improvement on the transition energy but not on the cross section, it has the following major advantage with respect to the original version of \textsc{xatom}:
With the help of the zeroth-order $LS$ eigenstates, the present implementation is capable to provide individual state-to-state cross sections and transition rates (see Sec.~\ref{CS} and~\ref{TR}), thus allowing us to study orbital alignment (see next section). 

\section{Results and discussion}\label{results}
We employ the individual state-to-state cross sections provided by our implementation to explore orbital alignment induced by linearly polarized x rays.
In particular, we consider the distribution of the states belonging to the ions produced by photoionization.
As a first example, we discuss photoionization of the neutral argon atom that is initially in a closed-shell configuration.
Having at hand the results for neutral argon, we then generalize them to some argon charge states in open-shell configurations. 
These ions can appear in the x-ray multiphoton ionization of neutral argon driven by an intense XFEL pulse.

Before starting, however, the following should be pointed out. 
In what follows we focus on the photoionization of an electron in a specific subshell of \(2p\) or \(3p\) ($l = 1$) without any interaction between the subshells (interchannel coupling). 
This is because (i) ionization of the subshell with \ \(l=0\) always completely aligns the remaining ion since only the final state with \({M}_{{L}_{f}}={M}_{{L}_{i}}\) is allowed and (ii) binding energies differ enough to neglect the interaction~\cite{Drake:IC}.
Furthermore, we focus on a specific initial zeroth-order $LS$ eigenstate. However, for \({M}_{L_i}\ne0\) we will consider a uniform distribution of initial states with \(\pm{M}_{L_i}\), denoted in the following by \(\vert{M}_{L_i}\vert\). This prevents an orientation of the final ion, which would be simply caused by a prior orientation of the initial ion. Therefore, the population probabilities of the final ion are identical  for \(\pm{M}_{{L}_{f}}\), i.e., \(p(-{M}_{L_f }\vert L_f) = p({M}_{L_f}\vert L_f)\). This is because Eqs.~(\ref{eqn:pcs}) to (\ref{eqn:amtr}) are identical for a transition from \({M}_{L_i}\) to \(\pm{M}_{L_f}\) and \(-{M}_{L_i}\) to \(\mp{M}_{L_f}\). So there is no orientation (see Sec.~\ref{A20}) and we can investigate the $\pm{M}_{{L}_{f}}$ cases together (without summing over both signs).

\subsection{Orbital alignment after ionization of neutral Ar}\label{Argon}
\begin{figure}[t]  
\centering
\includegraphics[trim = 0mm 65mm 0mm 0mm, clip, width=\linewidth]{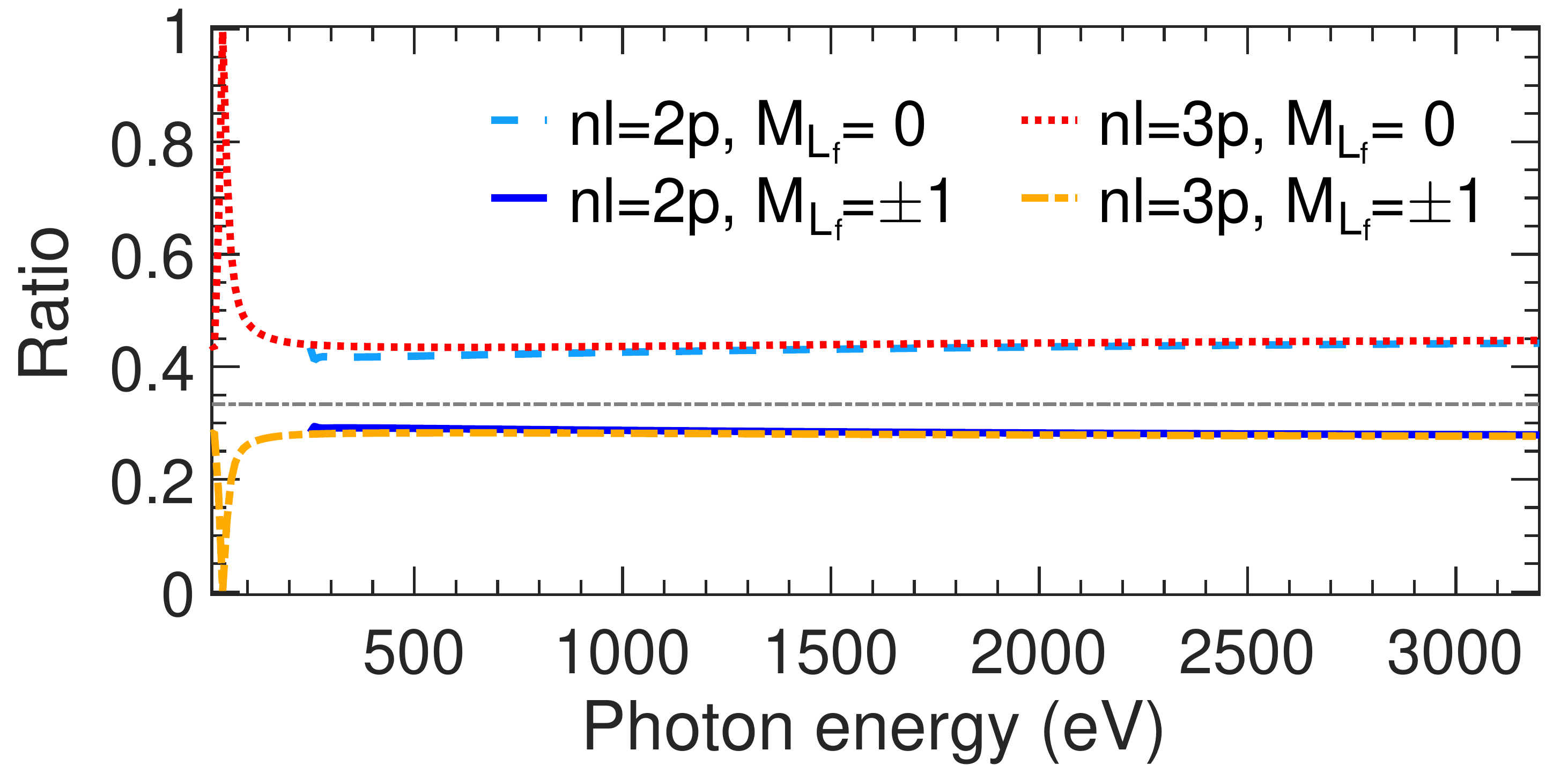}
\caption{Ratio of individual cross sections \({\sigma}_{^1\text{S};^2\text{P}}^{0;{M}_{{L}_{f}}}/{\sigma}_{^1\text{S}}^{0}\) for  neutral argon as a function of the photon energy  from the \(3p\) threshold  (\(\approx13.46~\text{eV}\)) to the \(1s\) threshold (\(\approx3207.51~\text{eV}\)). Results for different \({M}_{{L}_{f}}\) for both the ionization of the \(2p\) and \(3p\) subshells are shown (see legend). The gray line at $1/3$ indicates the case of a uniform distribution of \({M}_{{L}_{f}}\). In all cases, the atom is initially in the $^1$S state with \({M}_{{L}_{i}}=0\) and the final \ion{Ar}{} ion is in one of  the $^2$P states.}
\label{fig:Argon}
\end{figure} 
We first investigate the orbital alignment of the \ion{Ar}{} ion following the photoionization of the \(2p\) or \(3p\) subshell of neutral argon (\(1s^22s^22p^63s^23p^6\)) by linearly polarized x rays. The calculated ratios of individual cross sections \({\sigma}_{^1\text{S};^2\text{P}}^{0;{M}_{{L}_{f}}}/{\sigma}_{^1\text{S}}^{0}\) [Eq.~(\ref{eqn:ratio})] are presented in Fig.~\ref{fig:Argon} for all possible \({M}_{{L}_{f}}\)  as a function of the photon energy. Additionally, calculated alignment parameters \(\mathcal{A}_{20}(\text{P})\) [Eq.~(\ref{eqn:A20})] are listed in Table~\ref{tab:A} for various photon energies. For initially neutral argon \(\mathcal{A}_{20}(\text{P})\) can be directly obtained from the ratios in Fig.~\ref{fig:Argon}, i.e., \(\mathcal{A}_{20}(\text{P})=\sqrt{2}\lbrack{\sigma}_{^1\text{S};^2\text{P}}^{0;1}/{\sigma}_{^1\text{S}}^{0}-{\sigma}_{^1\text{S};^2\text{P}}^{0;0}/{\sigma}_{^1\text{S}}^{0}\rbrack\).
As it can be seen, in the x-ray regime, where the photon energy is greater than $\sim$300~eV, the resulting \ion{Ar}{} ion ($2p^{-1}$ or $3p^{-1}$) exhibits a clear orbital alignment (i.e., \(\mathcal{A}_{20} < 0\)), but it is not an extremely strong orbital alignment (i.e., \(\mathcal{A}_{20}\) is close to zero).
It increases only marginally with the photon energy and is a little stronger for the \(3p\) subshell than for the \(2p\) subshell. 
As a consequence of the alignment, almost \(45\%\) of the \ion{Ar}{} ions produced have an angular momentum projection of \({M}_{{L}_{f}}=0\), while the others have \({M}_{{L}_{f}}=\pm1\) with a probability of \(28\%\) for each \({M}_{{L}_{f}}\) (see Fig.~\ref{fig:Argon}). 
\begin{table}[b]
\centering
\caption{Alignment parameter \(\mathcal{A}_{20}(\text{P})\) of the final \ion{Ar}{} ion after ionization of neutral Ar. Results for both the ionization of the \(2p\) and \(3p\) subshell of neutral argon are listed for various photon energies. For comparison, note that a complete alignment with respect to \({M}_{L_f}=0\) would yield a value of \(\mathcal{A}_{20}(\text{P})=-\sqrt{2}\).}
\label{tab:A}
\begin{ruledtabular}
\begin{tabular}{crr}
\(nl\)&\({\omega}_{\text{in}}\) (eV)&\(\mathcal{A}_{20}(\text{P})\) \\
 \midrule
 \(3p\)&\(40\)&\(-1.406\)\\[4pt]
 \(2p\)&\(300\)&\(-0.178\)\\
 \(3p\)&\(300\)&\(-0.221\)\\[4pt]
  \(2p\)&\(1000\)&\(-0.196\)\\
 \(3p\)&\(1000\)&\(-0.218\)\\[4pt]
 \(2p\)&\(3000\)&\(-0.228\)\\
 \(3p\)&\(3000\)&\(-0.240\)\\
\end{tabular}
\end{ruledtabular}
\end{table}
\begin{table}[b]
\centering
\caption{Radial integrals \(\vert{R}_{{\varepsilon}_{c}0;{n}1}\vert\) and \(\vert{R}_{{\varepsilon}_{c}2;{n}1}\vert\) as well as their ratio for various photon energies for  neutral argon.}
\label{tab:R}
\begin{ruledtabular}
\begin{tabular}{crrrr}
\(nl\)&\({\omega}_{\text{in}}\) (eV)&\(~A=\vert{R}_{{\varepsilon}_{c}0;{n}1}\vert~\)&\(~B=\vert{R}_{{\varepsilon}_{c}2;{n}1}\vert\)& $\left( A / B \right)^2$ \\
 \midrule
 \(3p\)&\(40\)&\(2.99\times10^{-1}\)&\(0.71\times10^{-1}\)&\(302.00\)\\[4pt]
 \(2p\)&\(300\)&\(0.37\times10^{-1}\)&\(1.51\times10^{-1}\)&\(0.06\)\\
 \(3p\)&\(300\)&\(1.40\times10^{-2}\)&\(3.84\times10^{-2}\)&\(0.13\)\\[4pt]
  \(2p\)&\(1000\)&\(0.54\times10^{-2}\)&\(1.81\times10^{-2}\)&\(0.09\)\\
 \(3p\)&\(1000\)&\(1.75\times10^{-3}\)&\(4.89\times10^{-3}\)&\(0.13\)\\[4pt]
 \(2p\)&\(3000\)&\(0.71\times10^{-3}\)&\(1.85\times10^{-3}\)&\(0.15\)\\
 \(3p\)&\(3000\)&\(2.18\times10^{-4}\)&\(5.34\times10^{-4}\)&\(0.17\)\\
\end{tabular}
\end{ruledtabular}
\end{table}

There are two explanations for the observed alignment.
First, 
due to the angular momentum coupling of the photoelectron and the involved final hole for incoming radiation linearly polarized along the \(z\) axis~\cite{Hertel:AMOP, X-rayspectroscopy}, the ejection of an electron with \({m}_{{l}_{j}}=\pm1\) is less likely than with \({m}_{{l}_{j}}=0\). Hence, a transition with \({M}_{{L}_{i}}-{M}_{{L}_{f}}={m}_{l_j}=\pm1\) is less likely than one with \({M}_{{L}_{i}}={M}_{{L}_{f}}\). In particular, ratios of individual cross sections differ by a value of \(0.1\). 
This value can be obtained from an explicit calculation of Clebsch-Gordan coefficients in Eq.~(\ref{eqn:pcs}) and is, evidently, independent of the photon energy. 
Second, the remaining alignment can be explained as follows. 
Owing to the selection rules for a dipole transition~\cite{Hertel:AMOP}, the photoelectron can have two possible angular momentum quantum numbers, \({l}_{c}={l}\pm1\).  
In the calculation of cross sections \({l}_{c}\) is summed over both [see Eq.~(\ref{eqn:pcs})]. 
However, for \({M}_{{L}_{i}}-{M}_{{L}_{f}}=\pm1\) and \({l}=1\), the former is forbidden as \({l}_{c}=0<\vert{M}_{{L}_{i}}-{M}_{{L}_{f}}\vert\), so only \({l}_{c}=2\) contributes to the cross section. 
Consequently, when \({M}_{{L}_{i}}=0\), the cross section for \({M}_{{L}_{f}}=\pm1\) (only \({l}_{c}=2\) contributes) is smaller than that for \({M}_{{L}_{f}}=0\) (both \({l}_{c}=0\) and \({l}_{c}=2\) contribute). This effect becomes larger as  the  \({l}_{c}=0\)  contribution of the photoelectron increases.
From Eq.~(\ref{eqn:pcs}), we can conclude that the amount of this reduction depends on the ratio of radial integrals \(\vert{R}_{{\varepsilon}_{c}0;{n}1}\vert^2/\vert{R}_{{\varepsilon}_{c}2;{n}1}\vert^2\), where 
\begin{equation}
{R}_{{\varepsilon}_{c}{l}_{c};{n}l}=\int_{0}^{\infty}dr{u}_{{\varepsilon}_{c} {l}_{c}}^{\ast}(r)r{u}_{{n} l}(r).
\end{equation}
For initial neutral argon, the ratios of radial integrals are shown in Table~\ref{tab:R}. 
In the x-ray regime ($\omega_\text{in} \gtrsim 300$~eV), the ratio of radial integrals is quite small. 
Therefore, cross sections for \({M}_{{L}_{f}}=\pm1\) are reduced by the ratio of radial integrals only a little and, thus, the alignment is not extremely strong as shown in Table~\ref{tab:A}.
Worthy of note is also that the marginal increase of the alignment with the photon energy can be attributed to the ratio of radial integrals as well.
In contrast, below the x-ray regime at roughly 40~eV an opposite situation can be discovered as shown in Fig.~\ref{fig:Argon}, Table~\ref{tab:A}, and Table~\ref{tab:R}. 
Photoionization of the \(3p\) subshell by a linearly polarized photon with around 40~eV predominantly produces an ion with \({M}_{{L}_{f}}=0\) (i.e., \(\mathcal{A}_{20}\sim-\sqrt{2}\)).

\subsection{Orbital alignment after $2p$ ionization of \ion{Ar}{}~($2p^{-1}$)}\label{Ar+}
\begin{figure}[t]  
\centering
\includegraphics[trim = 0mm 65mm 0mm 0mm, clip, width=\linewidth]{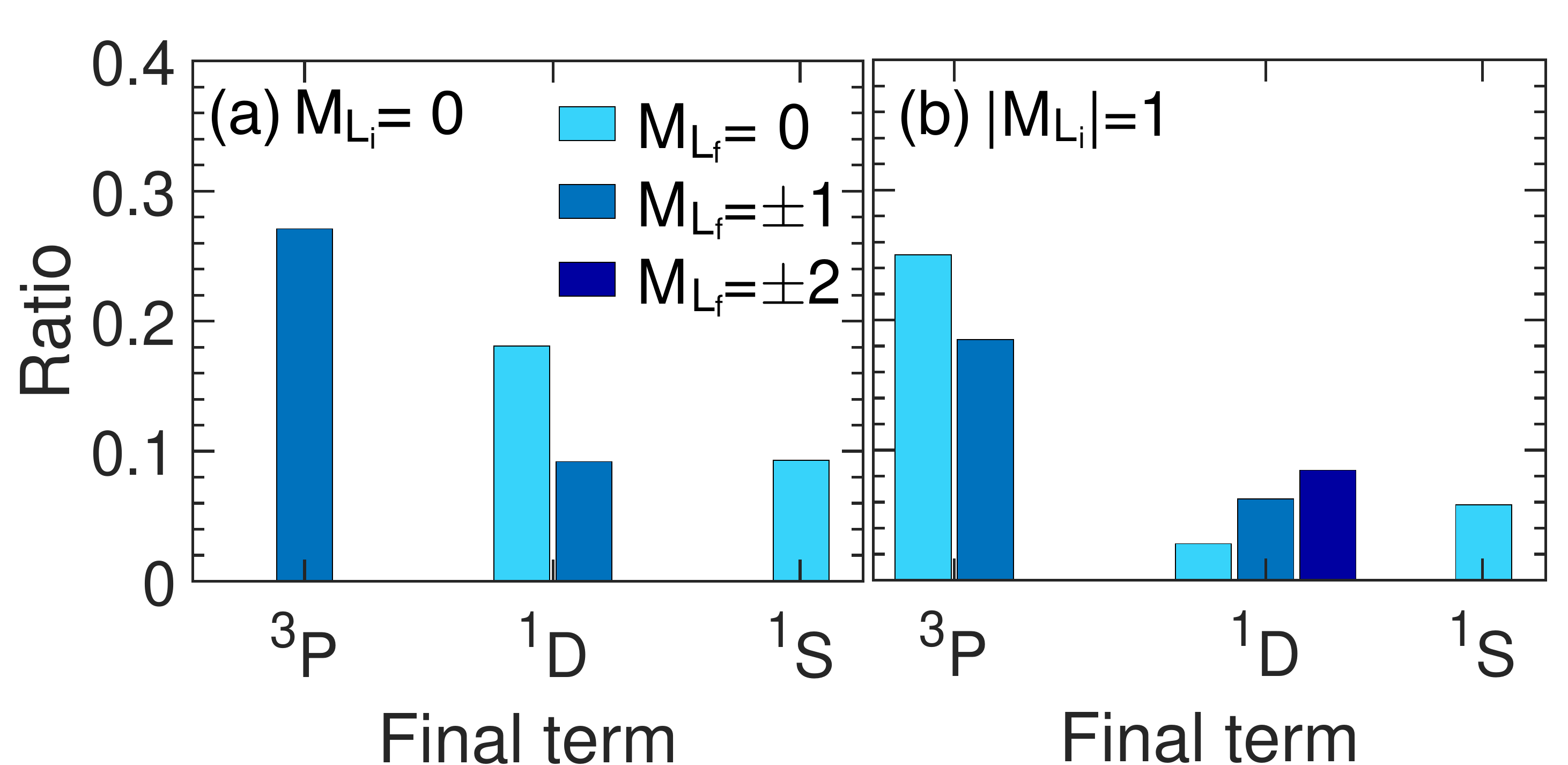}
\caption{Ratio of individual cross sections \({\sigma}_{^2\text{P};^{2S_f+1}L_f}^{\vert{M}_{{L}_{i}}\vert;{M}_{{L}_{f}}}/{\sigma}_{^2\text{P}}^{\vert{M}_{{L}_{i}}\vert}\) for \ion{Ar}{}~($2p^{-1}$) at a photon energy of \(1000~\text{eV}\) (a) for \({M}_{{L}_{i}}=0\) and (b) for \(\vert{M}_{{L}_{i}}\vert=1\) (i.e., uniform distribution of \({M}_{L_i}=\pm1\)). Results for different final terms and different \({M}_{{L}_{f}}\) are shown. In all cases, the atom initially is in a $^2$P state and the subshell being ionized is \(2p\). }\label{fig:Argonplus}
\end{figure} 
\begin{table}[b]
\centering
\caption{Alignment parameter \(\mathcal{A}_{20}(\text{P})\)  and  \(\mathcal{A}_{20}(\text{D})\) of the final \ion{Ar}{2} ion~($2p^{-2}$) after \(2p\) ionization of \ion{Ar}{} (\(2p^{-1}\) \(^2\)P). Results for different \(\vert{M}_{L_i}\vert\)  are listed at a  photon energy of \(1000~\text{eV}\) and \(3000~\text{eV}\). }
\label{tab:A+}
\begin{ruledtabular}
\begin{tabular}{crrr}
\({\omega}_{\text{in}}\) (eV)&\(\vert{M}_{L_i}\vert\)&\(\mathcal{A}_{20}(\text{P})\)&\(\mathcal{A}_{20}(\text{D})\) \\
 \midrule
\(1000\)& \(0\)&\(0.707\)&\(-0.895\)\\
\(1000\)& \(1\)&\(-0.147\)&\(0.290\)\\[4pt]
\(3000\)& \(0\)&\(0.707\)&\(-0.904\)\\
\(3000\)& \(1\)&\(-0.115\)&\(0.268\)\\
\end{tabular}
\end{ruledtabular}
\end{table}
Next we proceed to investigate orbital alignment after ionization of the initially open-shell configuration of \ion{Ar}{}~(\(1s^22s^22p^53s^23p^6\)). 
Here we are considering only $2p^{-1}$ for \ion{Ar}{} because (i) the partial cross section of $3p$ of neutral Ar is much smaller than that of $2p$, when the photon energy is greater than the $2p$ threshold and (ii) fluorescence processes that can also produce a hole in the \(3p\) subshell are very slow ($\sim2000$~fs lifetime) compared to the pulse durations of XFELs (a few fs).
Therefore, \ion{Ar}{} ions are barely found in the configuration \(1s^22s^22p^63s^23p^5\) during interaction with XFEL pulses. 
Likewise, we will focus on $2p$ ionization of \ion{Ar}{}~($2p^{-1}$) in the following discussion, because of the low cross section of the $3p$ subshell.

To explore the orbital alignment of the final \ion{Ar}{2}~($2p^{-2}$) that is produced by the photoionization of the \(2p\) subshell of \ion{Ar}{}, we calculate the alignment parameters  \(\mathcal{A}_{20}(\text{P})\)  and  \(\mathcal{A}_{20}(\text{D})\) as well as the ratios of individual cross sections \({\sigma}_{^2\text{P};^{2S_f+1}L_f}^{\vert{M}_{L_i}\vert;{M}_{{L}_{f}}}/{\sigma}_{^2\text{P}}^{\vert{M}_{L_i}\vert}\), i.e., \(\frac{1}{2}\lbrack{\sigma}_{^2\text{P};^{2S_f+1}L_f}^{+{M}_{{L}_{i}};{M}_{{L}_{f}}}/{\sigma}_{^2\text{P}}^{+{M}_{{L}_{i}}}+{\sigma}_{^2\text{P};^{2S_f+1}L_f}^{-{M}_{{L}_{i}};{M}_{{L}_{f}}}/{\sigma}_{^2\text{P}}^{-{M}_{{L}_{i}}}\rbrack\) for \({M}_{L_i}\ne0\). 
As observed above for neutral argon, we expect for the orbital alignment of initial \ion{Ar}{} ions a similar, very small and smooth change with the energy of the x-ray photons.  
The alignment parameters are listed at a photon energy of 1000 eV and 3000 eV in Table~\ref{tab:A+}. \(A_{20}(\text{P})\) and \(A_{20}(\text{D})\) are similar for both photon energies.  For this reason, we restrict ourselves here to the analysis of a photon energy  of \(1000~\text{eV}\) only in Fig.~\ref{fig:Argonplus}, where the ratio of individual cross sections is depicted for all possible initial states. 
The sum of the bars for each panel in Fig.~\ref{fig:Argonplus}, after taking into account a factor of 2 for $\pm M_{L_f}$, is equal to one. Note that the alignment parameters in Table~\ref{tab:A+} can be obtained from the relation between the bars belonging to a final term.

Combining the findings in Fig.~\ref{fig:Argonplus}(a) and Table~\ref{tab:A+}, we observe for \({M}_{{L}_{i}}=0\) that the majority of \ion{Ar}{2} ions produced are in the $^3$P state ($\sim54$\% uniformly distributed between \({M}_{{L}_{f}}=+1\) and \({M}_{{L}_{f}}=-1\)), which is completely aligned [i.e., \(\mathcal{A}_{20}(\text{P})=1/\sqrt{2}\)]. 
This alignment is simply caused by the selection rules for photoionization, where a transition to \({M}_{{L}_{f}}=0\) is forbidden. 
In particular, if \({L}_{i}+{L}_{f}+{l}\) is an odd number and \({M}_{{L}_{i}}=0\), then only \({M}_{{L}_{f}}\ne0\) is allowed. 
Next, most probably, is the production of ions in the $^1$D state ($\sim37$\%), whereas there is only a probability of less than \(10\%\) to find the final ion in the unaligned $^1$S state [always \(\mathcal{A}_{20}(\text{S})=0\)]. 
Note that also within the $^1$D term, there is orbital alignment [i.e., \(\mathcal{A}_{20}(\text{D})<0\)]: It is twice as likely to find the state with \({M}_{{L}_{f}}=0\) than those with  \({M}_{{L}_{f}}=\pm1\), and \({M}_{L_f}=\pm2\) is forbidden by the selection rules. Thus, it is very likely to observe an alignment of the final ion, either in a \(^3\text{P}\) or \(^1\text{D}\) state. 

Let us now discuss the outcome for \(\vert{M}_{{L}_{i}}\vert=1\) in Fig.~\ref{fig:Argonplus}(b) and Table~\ref{tab:A+}. 
Again the majority of \ion{Ar}{2} ions produced are in one of the $^3$P states ($\sim$62\%). 
However, in contrast to \({M}_{L_i} = 0\) here the \(^3\text{P}\) states exhibit only a weak alignment that is comparable to that for the residual \ion{Ar}{} discussed in Sec.~\ref{Argon}, but a little weaker.
Worthy of note is also the alignment of  the final $^1$D state attributed to a larger population of states with \({M}_{{L}_{f}}=\pm2\) than for smaller \({M}_{{L}_{f}}\) [for comparison a complete alignment with respect to \({M}_{L_f}=\pm2\) has \(\mathcal{A}_{20}(\text{D})=2\sqrt{5/14}\)].

Although most of the observations are related to angular momentum coupling, it is worthwhile to explain them explicitly with respect to the formula of the individual cross section [Eq.~(\ref{eqn:pcs})]. A detailed understanding might be useful in a future study of alignment during multiphoton ionization, which cannot be simply explained analytically by angular momentum coupling. The observations can be explained by a combination of the subsequent aspects.

First, in the computation of individual cross sections [Eq.~(\ref{eqn:pcs})], we sum over all accessible final spin projections, i.e., \({M}_{{S}_{f}}\). Owing to the selection rules, i.e., \(\vert{M}_{{S}_{i}}-{M}_{{S}_{f}}\vert\overset{!}{=}\frac{1}{2}\), for the final terms with \({S}_{f}={S}_{i}+\frac{1}{2}\) two final states (\({M}_{{S}_{f}}={S}_{i}\pm\frac{1}{2}\)) are involved in the transition. 
In contrast, for the final terms with \({S}_{f}={S}_{i}-\frac{1}{2}\) only one final state (\({M}_{{S}_{f}}={S}_{i}-\frac{1}{2}\)) is allowed. 
As a consequence, the states for the former terms tend to have higher cross sections. 
It is because of the cross section being independent of the initial spin projection that this argument is true for general initial states. 
Therefore, we conclude that final states with \({S}_{f}={S}_{i}+\frac{1}{2}\) are generally more probable than those with \({S}_{f}={S}_{i}-\frac{1}{2}\). 
This explains why transitions to the final $^3$P states of \ion{Ar}{2} are so dominant.

Second, if not forbidden, transitions preserving the angular momentum projection, i.e., \({M}_{{L}_{f}}={M}_{{L}_{i}}\), are generally preferred, while those changing it by one are suppressed. 
This is attributed to the fact that the incoming x rays are linearly polarized (see Sec.~\ref{Argon}). 
It explains, for instance, why final $^1$S states are a little more likely for \({M}_{{L}_{i}}=0\) than for \({M}_{{L}_{i}}=\pm1\). It also explains the alignment  of  the $^3$P states  for \(\vert{M}_{{L}_{i}}\vert=1\) and that of the $^1$D states for \({M}_{{L}_{i}}=0\). 

However, it does not explain the alignment of the $^1$D states for \(\vert{M}_{{L}_{i}}\vert=1\) or more precisely why the states with \({M}_{{L}_{f}}=\pm2\) are more probable than the other $^1$D states.
This can be explained by the square of the overlap matrix element, \(\braketoperator{{L}_{f}{S}_{f}{M}_{{L}_{f}}{M}_{{S}_{f}}}{\hat{c}_{j}}{{L}_{i}{S}_{i}{M}_{{L}_{i}}{S}_{i}}\), contained also in the formula for the individual cross section in Eq.~(\ref{eqn:pcs}). 
Thus, the third point is that this overlap matrix element additionally affects the orbital alignment.  
Obviously, a transition with a higher overlap matrix element, is more likely than that with a smaller overlap matrix element. 
Hence, the ratio of individual cross sections corresponding to the transition, being more probable, is enhanced. 
Therefore, transitions that do not preserve the angular momentum projection can be preferred when they exhibit very high overlap matrix elements. 
Regarding the final term $^1$D, the final states with \({M}_{{L}_{f}}=\pm2\) are pure Fock states and so is the initial $^2$P state. 
When the transition is not forbidden, for pure Fock states, the matrix element \(\braketoperator{{L}_{f}{S}_{f}{M}_{{L}_{f}}{M}_{{S}_{f}}}{\hat{c}_{j}}{{L}_{i}{S}_{i}{M}_{{L}_{i}}{S}_{i}}\) is evidently unity (maximal possible value). 
Thus, this transition is quite dominant.
On the other hand, the alignment related to the previously-mentioned point can be enhanced, when overlap matrix elements are higher for transitions with \({M}_{{L}_{f}}={M}_{{L}_{i}}\) than for the others. 
Note that this is the case for a transition from the initial state with \({M}_{{L}_{i}}=0\) to the final $^1$D states. 
In this context it is worth mentioning that for neutral argon (Sec.~\ref{Argon})  the overlap matrix element is always unity, because all involved states are pure Fock states.

Finally, some transitions are directly forbidden by the selection rules for photoionization (given in the middle of this section).
This is another important, but trivial reason for the orbital alignment.

\subsection{Orbital alignment after $2p$ ionization of \ion{Ar}{2}~($2p^{-2}$)}\label{Ar2+}
\begin{figure}[t]  
\centering
\includegraphics[width=\linewidth]{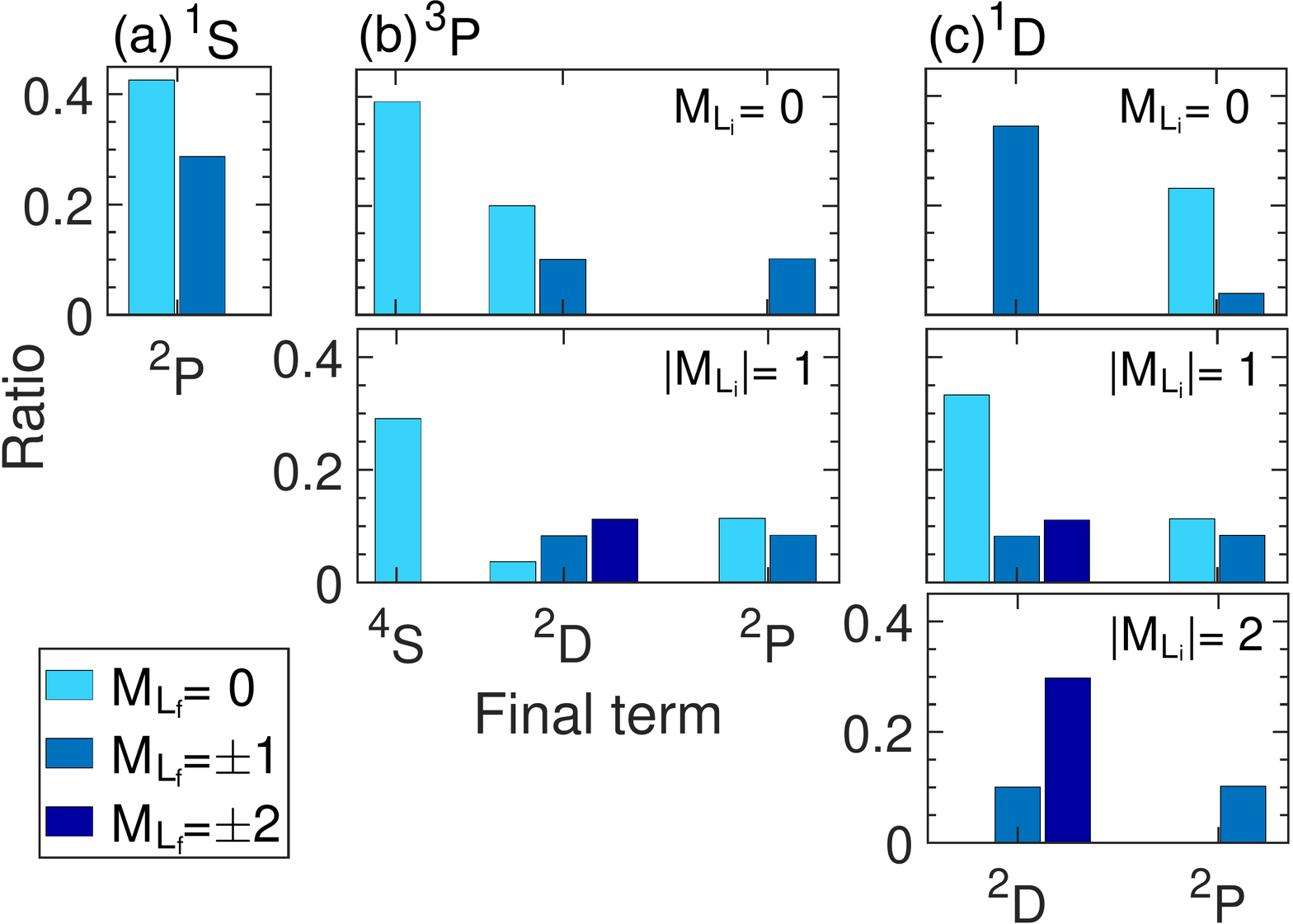}
\caption{Ratio of individual cross sections \({\sigma}_{^{2S_i+1}L_i;^{2S_f+1}L_f}^{\vert{M}_{{L}_{i}}\vert;{M}_{{L}_{f}}}/{\sigma}_{^{2S_i+1}L_i}^{\vert{M}_{{L}_{i}}\vert}\) for \ion{Ar}{2}~($2p^{-2}$) at a photon energy of \(1000~\text{eV}\). The atom is initially in (a) the $^1$S state with \({M}_{L_i}=0\), (b) one of the $^3$P states, or (c) one of the $^1$D states. Results for different \(\vert{M}_{{L}_{i}}\vert\) (i.e., uniform distribution of \(\pm{M}_{L_i}\)), different final terms, and different \({M}_{{L}_{f}}\)  are shown. In all cases, the subshell being ionized is \(2p\).}
\label{fig:Argonplus2}
\end{figure}
\begin{table}[b]
\centering
\caption{Alignment parameter \(\mathcal{A}_{20}(\text{P})\)  and  \(\mathcal{A}_{20}(\text{D})\) of the final \ion{Ar}{3} ion~($2p^{-3}$) after \(2p\) ionization of \ion{Ar}{2} (\(2p^{-2}\)). Results for different  initial states   are listed at a  photon energy of \(1000~\text{eV}\) and \(3000~\text{eV}\).}
\label{tab:A++}
\begin{ruledtabular}
\begin{tabular}{crrrr}
\({\omega}_{\text{in}}\) (eV)&\(^{2S_i+1}L_i\)&\(\vert{M}_{L_i}\vert\)&\(\mathcal{A}_{20}(\text{P})\)&\(\mathcal{A}_{20}(\text{D})\) \\
 \midrule
\(1000\)&\(^{3}\text{P}\)&\(0\)&\(0.707\)&\(-0.895\)\\
\(1000\)&\(^{3}\text{P}\)&\(1\)&\(-0.148\)&\(0.290\)\\[4pt]
\(1000\)&\(^{1}\text{D}\)&\(0\)&\(-0.879\)&\(-0.598\)\\
\(1000\)&\(^{1}\text{D}\)&\(1\)&\(-0.148\)&\(-0.321\)\\
\(1000\)&\(^{1}\text{D}\)&\(2\)&\(0.707\)&\(0.743\)\\[4pt]
\(1000\)&\(^{1}\text{S}\)&\(0\)&\(-0.196\)&\\[4pt]
\(3000\)&\(^{3}\text{P}\)&\(0\)&\(0.707\)&\(-0.904\)\\
\(3000\)&\(^{3}\text{P}\)&\(1\)&\(-0.114\)&\(0.267\)\\[4pt]
\(3000\)&\(^{1}\text{D}\)&\(0\)&\(-0.905\)&\(-0.598\)\\
\(3000\)&\(^{1}\text{D}\)&\(1\)&\(-0.114\)&\(-0.325\)\\
\(3000\)&\(^{1}\text{D}\)&\(2\)&\(0.707\)&\(0.765\)\\[4pt]
\(3000\)&\(^{1}\text{S}\)&\(0\)&\(-0.230\)&\\
\end{tabular}
\end{ruledtabular}
\end{table}
In order to complete our understanding of orbital alignment we finally investigate photoionization of \ion{Ar}{2}~(\(1s^22s^22p^43s^23p^6\)) as another example of an initial open-shell configuration. 
For the reasons explained in Sec.~\ref{Ar+}, the focus is again on the photoionization of the \(2p\) subshell.
To characterize the orbital alignment of the final \ion{Ar}{3}~($2p^{-3}$), we show calculated alignment parameters \(\mathcal{A}_{20}(\text{P})\) and \(\mathcal{A}_{20}(\text{D})\) in Table~\ref{tab:A++} (at photon energies of 1000 eV and 3000 eV) and ratios of individual cross sections \({\sigma}_{^{2S_i+1}L_i;^{2S_f+1}L_f}^{\vert{M}_{{L}_{i}}\vert;{M}_{{L}_{f}}}/{\sigma}_{^{2S_i+1}L_i}^{\vert{M}_{{L}_{i}}\vert}\) in Fig.~\ref{fig:Argonplus2} (at 1000 eV only). 
It becomes evident that the degree of alignment for the \ion{Ar}{3} ions produced  is comparable with that observed in the previous cases. 
Above all, the observations for the final \ion{Ar}{3} ions can be explained by the same arguments as provided for the final \ion{Ar}{2} ions in Sec.~\ref{Ar+}. 

Nonetheless, three things should be pointed out.
First, for the initial $^3$P states of \ion{Ar}{2}, the most probable final state of \ion{Ar}{3} is the unaligned $^4$S state ($\sim$40\% for \({M}_{{L}_{i}}=0\) and $\sim$30\% for \(\vert{M}_{{L}_{i}}\vert=1\)), as shown in Fig.~\ref{fig:Argonplus2}(b). 
As argued in the first point in Sec.~\ref{Ar+}, it is because of the spin quantum number being the highest for the final $^4$S state that this state becomes dominant here.
Second, for the initial $^1$D state with \(\vert{M}_{{L}_{i}}\vert=2\), more than half of the resulting \ion{Ar}{3} ions retain the angular momentum quantum number and its projection, i.e., \({L}_{f}=2\) and \({M}_{{L}_{f}}=\pm2\), as depicted in Fig.~\ref{fig:Argonplus2}(c). This leads to a comparably strong alignment of the $^2$D states [see Table~\ref{tab:A++} and compare with \(\mathcal{A}_{20}(\text{D})=2\sqrt{5/14}\) for a complete alignment with respect to \({M}_{L_f}=\pm2\)].
This alignment stems from the facts that \({M}_{{L}_{f}}={M}_{{L}_{i}}\) transitions are preferred (Sec.~\ref{Argon}) and the overlap matrix element takes on the highest possible value for pure Fock states (see the third point in Sec.~\ref{Ar+}).  
Third, note also the comparably high alignment of the $^2$P state for initial $^1$D states with \({M}_{L_i}=0\) [compare with \(\mathcal{A}_{20}(\text{P})=-\sqrt{2}\) for a complete alignment with respect to \({M}_{L_f}=0\)]. This can be explained by the same arguments as provided for the $^2$D states. In particular, here the overlap matrix element for \({M}_{L_f}=0\) is two times that for \({M}_{L_f}=\pm1\).
Another important remark here is that for initial states with equal \(L_i\) and \({M}_{L_i}\) the alignment of the final states is almost independent of the charge state of argon and of the spin multiplicity of the initial and final states. This can be seen by comparing the alignment parameters given in Tables~\ref{tab:A}, \ref{tab:A+}, and \ref{tab:A++}.
All this is closely related to the fact that alignment mainly depends on angular momentum coupling and that ratios of radial integrals, which additionally affect the alignment, are very similar  for all charge states of argon (not shown here for brevity).

\subsection{Orbital alignment after Auger-Meitner decay of \ion{Ar}{} (\(2p^{-1}\))}\label{AM}
\begin{figure}[t]  
\centering
\includegraphics[trim = 10mm 65mm 10mm 0mm, clip, width=\linewidth]{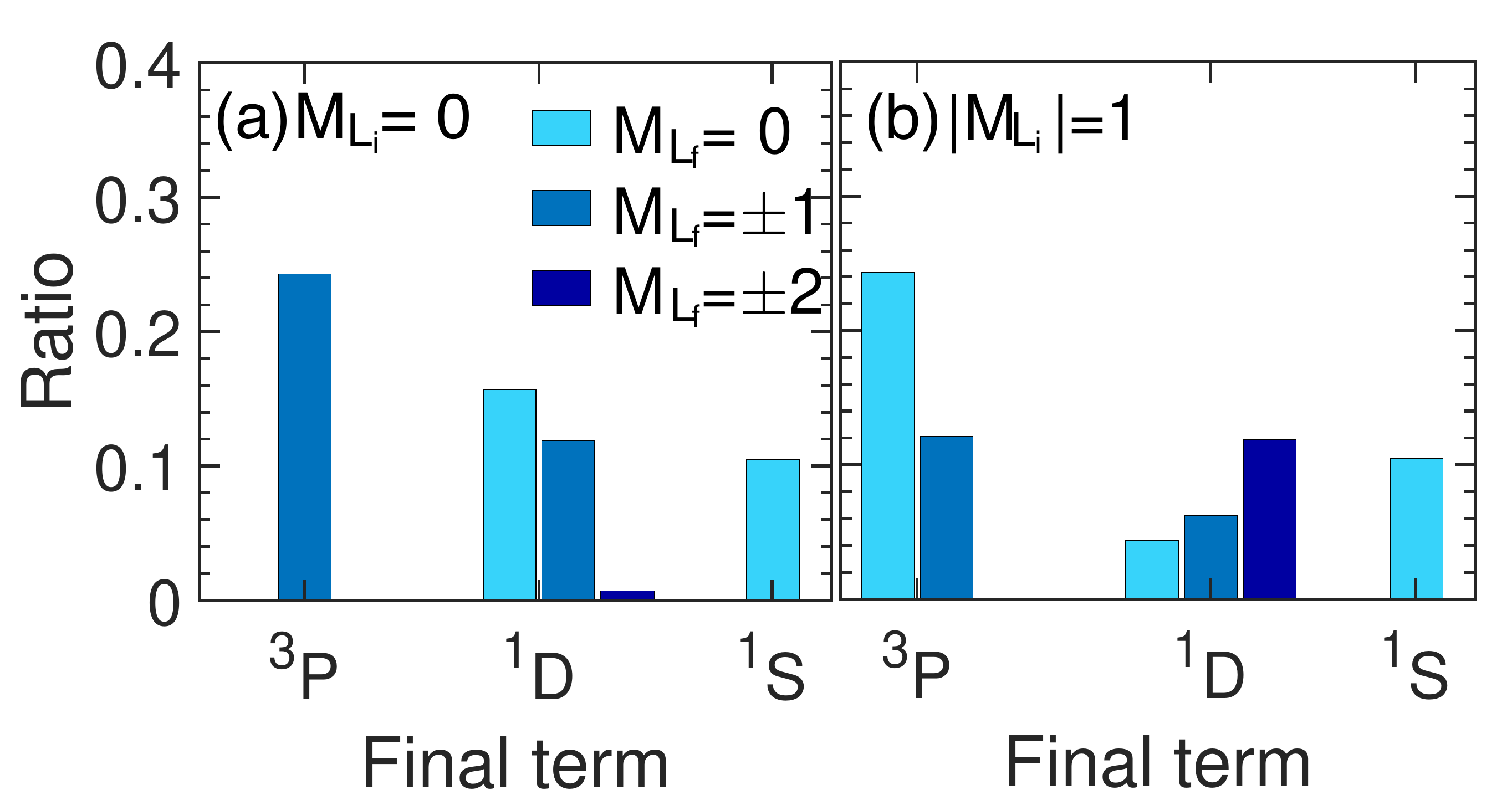}
\includegraphics[trim = 10mm 65mm 10mm 0mm, clip, width=\linewidth]{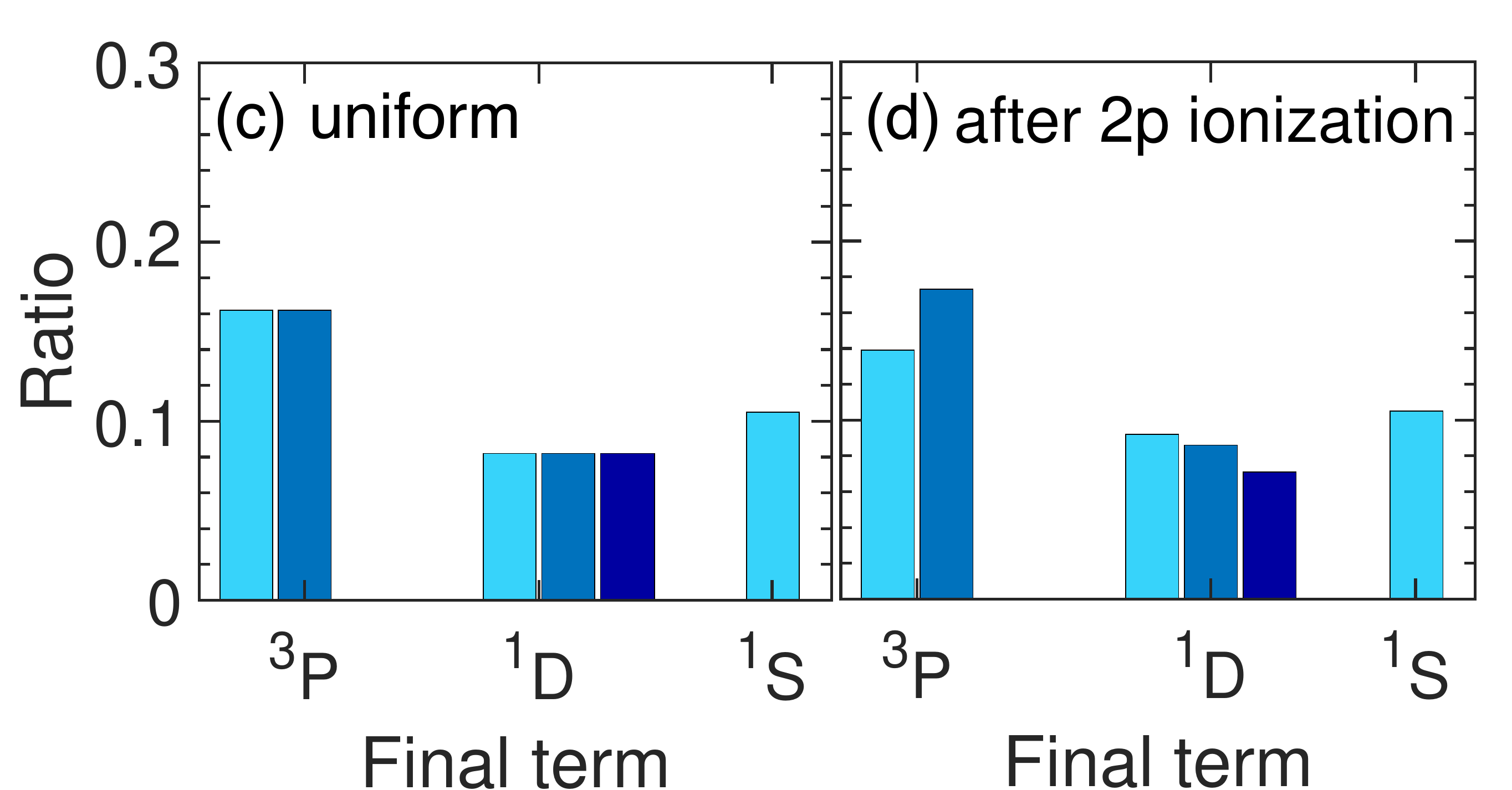}
\caption{Ratio of individual Auger-Meitner transition rates \({\Gamma}_{^2\text{P};^{2S_f+1}L_f}^{\vert{M}_{{L}_{i}}\vert;{M}_{{L}_{f}}}/{\Gamma}_{^2\text{P}}^{\vert{M}_{{L}_{i}}\vert}\) for \ion{Ar}{}~($2p^{-1}$)  (a) for \({M}_{{L}_{i}}=0\) and (b) for \(\vert{M}_{{L}_{i}}\vert=1\) (i.e., uniform distribution of \({M}_{L_i}=\pm1\)). The weighted mean of the ratios with respect to \(M_{L_i}\) is shown in (c) for a uniform distribution of \(M_{L_i}\)  and in (d) for the distribution of \(M_{L_i}\) after \(2p\) ionization of neutral argon at 1000 eV.  Results for different final terms and different \({M}_{{L}_{f}}\) are shown. In all cases, the atom initially is in a $^2$P state and decays via \(L_{23}M_{23}M_{23}\) Auger-Meitner decay. }\label{fig:barplotAM}
\end{figure} 
\begin{table}[t]
\centering
\caption{Alignment parameter \(\mathcal{A}_{20}(\text{P})\)  and  \(\mathcal{A}_{20}(\text{D})\) of the final \ion{Ar}{2} ion~($3p^{-2}$) after \(L_{23}M_{23}M_{23}\) Auger-Meitner decay of \ion{Ar}{} (\(2p^{-1}\) \(^2\)P). Results for different \(\vert{M}_{L_i}\vert\)  are listed.}
\label{tab:AMdecay}
\begin{ruledtabular}
\begin{tabular}{crr}
\(\vert{M}_{L_i}\vert\)&\(\mathcal{A}_{20}(\text{P})\)&\(\mathcal{A}_{20}(\text{D})\) \\
 \midrule
 \(0\)&\(0.707\)&\(-0.770\)\\
 \(1\)&\(-0.354\)&\(0.385\)\\
\end{tabular}
\end{ruledtabular}
\end{table}
We would like to point out that the lifetime of the \(2p\) hole in \ion{Ar}{} (\(2p^{-1}\)) is about 3.9~fs, that of the \(2p^2\) hole in \ion{Ar}{2} (\(2p^{-2}\)) is about 1.6~fs, and that of the \(2p^3\) hole in \ion{Ar}{3} (\(2p^{-3}\)) is only about 0.9~fs  (calculated with the first-order strategy). 
Therefore, it is likely that the transient hole states produced by photoionization undergo an Auger-Meitner decay before further photoionization can occur, unless the x-ray intensity is extremely high. Note that  decay also happens via fluorescence, but fluorescence rates are much smaller than Auger-Meitner decay rates for the argon ions.
As a consequence, it is indispensable to involve Auger-Meitner decay processes in the studies of orbital alignment. This becomes especially important when 
investigating orbital alignment dynamics of ions produced by x-ray multiphoton ionization.

Here we consider as an example the \(L_{23}M_{23}M_{23}\) Auger-Meitner decay of \ion{Ar}{} (\(2p^{-1}\)), so the final configuration is \ion{Ar}{2} (\(3p^{-2}\)).
 To explore the orbital alignment of the final \ion{Ar}{2} (\(3p^{-2}\)),  we  calculate alignment parameters \(\mathcal{A}_{20}(\text{P})\)  and  \(\mathcal{A}_{20}(\text{D})\) via Eq.~(\ref{eqn:A20}) in Table~\ref{tab:AMdecay} and ratios of individual transition rates via Eq.~(\ref{eqn:amtr}) in Figs.~\ref{fig:barplotAM}(a) and \ref{fig:barplotAM}(b). As can be seen, for a fixed initial state, i.e., only one \(\vert M_{L_i} \vert\), Auger-Meitner decay leads to a clear alignment of the final  \ion{Ar}{2} ion. However, if the initial state has no alignment, i.e., a uniform distribution of \(M_{L_i}\), then the weighted means of the ratios of individual transition rates are equal for all \({M}_{L_f}\) belonging to a  final term [see Fig.~\ref{fig:barplotAM}(c)]. 
Here, the weighted mean is the sum over the ratios of individual transition rates for all possible \(M_{L_i}\), weighted by the population probability \(p({M}_{L_i}\vert L_i)\) of the initial state. Note that in the uniform case \(p({M}_{L_i}\vert L_i)\) equals \(1 / (2L_i+1)\).
Consequently, the final  \ion{Ar}{2} ion produced by Auger-Meitner decay of an unaligned \ion{Ar}{} ion does not possess any alignment. We have \(\mathcal{A}_{20}(\text{P})=0\)  and  \(\mathcal{A}_{20}(\text{D})=0\) when taking the weighted mean of the alignment parameters given in Table~\ref{tab:AMdecay}, a factor of 2 for \(\vert M_{L_i}\vert = 1\) included. 
The reason for the zero alignment is the following. According to the Wigner-Eckhart theorem~\cite{Judd:OTiAS,Racah3} the transition rate is independent of the angular momentum projection of the total initial and final electronic state, the Auger electron included. Thus, a sum over the individual transition rates in Eq.~(\ref{eqn:amtr}) is independent of the final projection \(M_{L_f}\) when it is uniformly summed over the initial projection \(M_{L_i}\).
Therefore, in general the Auger-Meitner decay processes will not create any alignment of the final ion if it initially starts with zero alignment, i.e., a uniform distribution.

However, if the initial ion is already aligned, then the final ion produced by Auger-Meitner decay will show an alignment as demonstrated in the next section.

\subsection{Orbital alignment evolution in x-ray-induced ionization process}\label{PA}
\begin{table}[t]
\centering
\caption{Alignment parameter \(\mathcal{A}_{20}(\text{P})\)  and  \(\mathcal{A}_{20}(\text{D})\) of the final \ion{Ar}{2}ion~($3p^{-2}$) after a sequence of \(2p\) ionization and \(L_{23}M_{23}M_{23}\) Auger-Meitner decay of initial neutral argon. Results for different photon energies are listed.}
\label{tab:AMsequence}
\begin{ruledtabular}
\begin{tabular}{crr}
\(\omega_{\text{in}} (eV)\)&\(\mathcal{A}_{20}(\text{P})\)&\(\mathcal{A}_{20}(\text{D})\) \\
 \midrule
300&\(0.089\)&\(-0.097\)\\
1000&\(0.098\)&\(-0.107\)\\
3000&\(0.114\)&\(-0.124\)\\
\end{tabular}
\end{ruledtabular}
\end{table}
Let us finally discuss how the alignment evolves in a sequence comprising a  photoionization event and an Auger-Meitner decay. We start with neutral argon (\(1s^22s^22p^63s^23p^6\)) and ionize the \(2p\) subshell as investigated in Sec.~\ref{Argon}. Then the produced \ion{Ar}{} ion (\(2p^{-1}\))  undergoes one \(L_{23}M_{23}M_{23}\) Auger-Meitner decay. For a fixed initial state, i.e., \(\vert M_{L_i}\vert\), and a uniform distribution the corresponding alignment has been investigated in the previous section. Here, the population probability \(p({M}_{L_i}\vert L_i)\) of \ion{Ar}{} (\(2p^{-1}\)) before the Auger-Meitner decay  is given by the ratios of individual cross sections for neutral argon in Fig.~\ref{fig:Argon}. To examine the alignment of the final \ion{Ar}{2} ion (\(3p^{-2}\)) after the sequence of \(2p\) photoionization and \(L_{23}M_{23}M_{23}\) Auger-Meitner decay, means of the ratios of individual transition rates are shown in Fig.~\ref{fig:barplotAM}(d) (at a photon energy of 1000 eV) and alignment parameters are shown in Table~\ref{tab:AMsequence} (for 300~eV, 1000~eV, and 3000~eV). Most importantly, we observe that the \ion{Ar}{2} ion exhibits a slight alignment, which is much smaller than that for \( M_{L_i} = 0\) (see Table~\ref{tab:AMdecay}). The reason for this is that the transiently \ion{Ar}{} ions produced by \(2p\) photoionization also possess only a weak alignment (see Fig.~\ref{fig:Argon} and Table~\ref{tab:A}). Thus, they are close to the uniform distribution with zero alignment (see Sec.~\ref{AM}).
For increasing photon energies this alignment after \(2p\) photoionization of neutral argon becomes a little stronger (see Fig.~\ref{fig:Argon} and Table~\ref{tab:A}) and with this also the alignment of \ion{Ar}{2} after the sequence of photoionization and Auger-Meitner decay (see Table~\ref{tab:AMsequence}).

\section{Conclusion}\label{conclusion}
In this paper, we have presented an implementation of improved electronic-structure calculations in the \textsc{xatom} toolkit that provides individual zeroth-order $LS$ eigenstates by employing first-order many-body perturbation theory. 
Based on this implementation, we have calculated individual state-to-state photoionization cross sections and transition rates. We have investigated orbital alignment after either single photoionization or one Auger-Meitner relaxation process, and then the evolution of orbital alignment in a sequence of photoionization and relaxation.

To set the stage, we have first presented a brief outline of the underlying method to calculate first-order-corrected energies and zeroth-order $LS$ eigenstates for arbitrary electronic configurations. 
We have also shown an analytical expression for the individual state-to-state cross section and transition rates.
Comparing \(K\alpha\) fluorescence energies and \(KLL\) Auger-Meitner electron energies of \ion{Ne}{}~($1s^{-1}$) with experimental data, we have confirmed that the extended \textsc{xatom} toolkit can describe transition energies significantly better than the original version. 
On the other hand, we have observed almost no improvement on the total photoionization cross sections for neutral argon, which can be attributed to the use of zeroth-order states.

Having the capability of calculating individual state-to-state cross sections by using the extended \textsc{xatom} toolkit, we have investigated orbital alignment induced by linearly polarized x rays for initial neutral argon and two exotic open-shell configurations of argon. 
Some degrees of alignment has been found for a wide range of x-ray photon energies. 
For initial neutral argon, the ions produced by photoionization exhibit a clear preference for conservation of the angular momentum projection.
For the initial open-shell ions, however, the distribution of final states is affected not only by the conservation of the angular momentum projection, but also by the final total spin quantum number, the selection rules, and, most importantly, the overlap matrix element. Finally, we have showcased how the orbital alignment is affected by Auger-Meitner decay and how it evolves during one sequence of photoionization and Auger-Meitner decay.

There are several promising perspectives for further developments.
Above all, the individual state-to-state cross sections and transition rates calculated with our implementation could be embedded in the rate-equation model employed in the \textsc{xatom} toolkit~\cite{IoIXBwA,Son, Son2012}. 
Solving rate equations would enable investigations of orbital alignment dynamics of ions produced by x-ray multiphoton ionization. 
In this way, it could be explored whether the orbital alignment observed here for ions produced by single photoionization is enhanced or reduced by successive   photoionization events and accompanying decay processes. 
Another interesting perspective is the improvement of the cross section by calculating and utilizing not only first-order-corrected energies but also first-order states and by taking interchannel coupling~\cite{IC} into account.
Lastly, relativistic effects and resonance effects are incorporated in the \textsc{xatom} toolkit~\cite{Toyota} but remains to be addressed in combination with the present implementation. Such methodological developments are not only important for many practical applications of focused XFEL beams, but are also useful for a quantitative characterization of XFEL beam properties.

%

\end{document}